\documentclass[preprint] {elsarticle} 
\journal{Journal of Applied Mechanics.} 
\usepackage{lineno,hyperref}
\pdfoutput=1 
\modulolinenumbers[5]

\biboptions{authoryear}

\usepackage[utf8]{inputenc} 

\usepackage{geometry} 
\usepackage{color}

\usepackage{amsmath}
\usepackage{amsfonts}
\usepackage{mathrsfs} 
\usepackage{booktabs} 
\usepackage{array} 
\usepackage{paralist} 
\usepackage{verbatim} 
\usepackage{subfig} 
\usepackage[color=green!40,disable]{todonotes}
\usepackage{versions}
\usepackage{bm}
\usepackage{url}

\usepackage{tikz}  
\usepackage{pgfplots}
\pgfplotsset{compat=1.17}

\newcommand\be{\begin{equation}}
\newcommand\ee{\end{equation}}

\newcommand{\mb}[1]{\mathbf{#1}}
\newcommand{\rr}[1]{\textcolor{red}{#1}}
\newcommand{\bb}[1]{\textcolor{blue}{#1}}

\begin{document}

\begin{frontmatter}

\title{Mechanics of curved crease origami: 1DoF mechanisms, distributed actuation by spontaneous curvature, and cross-talk between multiple folds}

\author[ADS1,ADS2]{Antonio DeSimone\corref{cor1}
}\ead{a.desimone@santannapisa.it, desimone@sissa.it}
\address[ADS1]{The BioRobotics Institute, Scuola Superiore Sant'Anna, viale Rinaldo Piaggio 34, 56025 Pontedera, Italy}
\address[ADS2]{SISSA--International School for Advanced Studies, via Bonomea 265, 34136 Trieste, Italy}
\author[LT]{Luciano Teresi}
\address[LT]{Department of Industrial Engineering--Universit\`a Roma TRE, 00146 Roma, Italy}
\cortext[cor1]{Corresponding author.}

\begin{abstract}
Origami morphing, obtained with patches of piecewise smooth isometries separated by straight fold lines, is an exquisite art that has already received considerable attention in the mathematics and mechanics literature. Curved fold lines, leading to curved creases and curved pleated structures, introduce the additional complexity of mechanical coupling between the folds.
This coupling can be exploited to obtain morphing structures with more robust folding pathways. We discuss one-degree-of-freedom mechanisms and folds actuated by spontaneous curvature (as in the case of hygromorphic multilayered composites),
comparing the purely geometric approach to two approaches based on the mechanics of active shells and of active three-dimensional solids. Moreover, we discuss the cooperativity of multiple folds and demonstrate the energetic advantage of synchronous folding over sequential folding.
\end{abstract}


\end{frontmatter}


%

\tableofcontents


\section{Introduction}

Origami morphing,  typically obtained with patches of piecewise smooth isometries separated by straight fold lines, is an exquisite art that has already received considerable attention in the mathematics and mechanics literature.

The variant in which the fold lines are not straight is particularly intriguing. Besides the interest acquired in the fields of art, design, manufacturing, and architecture, mastering curved pleated structures requires answering to a number of fundamental questions. From the point of view of mechanics, curvature of the fold-lines introduces the opportunities and challenges due to the fact that the folds become 
mechanically coupled. This can be exploited to engineer complex deployable structures whose folding and unfolding can  nevertheless be controlled robustly (i.e., in a foolproof fashion) through the actuation of few degrees of freedom (DoFs): we discuss here an important example of a 1 DoF mechanism and of its variant obtained by diffuse actuation via spontaneous curvature. In addition, coupling between the folds can be exploited to engineer morphing structures exhibiting multi-stability and preferred folding pathways: as an example, we discuss synchronous versus sequential folding for hyper-extensible tubular structures.

The differential geometry of origami obtained by folding along curves, rather than straight lines, has also been investigated by the mathematical community. While the local geometry of folding along a single curve is well understood (see Section \ref{sec:geometric_approach} and Appendix 1), unraveling the global consequences of nontrivial patterns of fold lines (such as, for example, the case of an annulus with multiple concentric circular
fold lines, see \cite{alese2022propagation}) poses challenges that are still unmet. The key problem here is to understand
the cross-talk between the various prescribed fold-lines. 
This question of how a curved fold propagates to the next prescribed fold line is very closely related to the one of how distinct neighboring folds interact mechanically, which is the central question motivating our study. 

In this paper, we focus on the interplay between the geometry and the mechanics of curved folds, and on the issue of cooperative vs antagonistic interactions among multiple folds.
Our approach has been inspired by the large existing literature, from which we single out the following contributions: \cite{fuchs_omnibus},
\cite{tachi2011one}, \cite{demaine2018conic}, \cite{jiang2019curve},
\cite{alese2022propagation}, \cite{gao2020shape}, \cite{tahouni2020self},
\cite{liu2024design}, \cite{feng2024geometry}.

\section{Two motivating examples: results using geometric models, active shells, and  3D active solids}

We set the stage for our discussion of curved-crease origami by first considering two key
examples.
The first is the analysis of a single fold line shaped as a circular arc and actuated by the folding angle in the center, 
see Section~\ref{sec:geometric_approach}. Here, the results are the proof that this is a 1 DoF mechanism and that there is a maximal extent of the fold line before the (local) construction breaks down because of overlapping of the two folds.

The second motivating example 
is the analysis of the tapering of two curved pleats separated by a single fold line shaped as a circular arc and
actuated by 
spontaneous (or target) curvature of the pleats, 
see Section~\ref{sec:elastic}.
By `tapering' we mean the increase of the visible curvature of the pleats towards the target value as we move away from the curved crease that joins them.
Here, the result is an explanation of the tapered shape predicted by mechanics and observed experimentally in synthetic hygromorphic structures.

The second example, where we consider a mechanism of diffuse actuation throughout the surface of the folding structure should be contrasted against the first one, where actuation is concentrated, through the control of the folding angle at a single point along the fold line. 

\subsection{Geometric approach: actuation by folding angle}\label{sec:geometric_approach}
In this section, we present the 
theory for the local geometry of folding along a single curve, and discuss some of its applications. We start by introducing the notation
to describe the geometry of the sheet and of the curved fold-line, both in the flat state before folding, and in the deformed state reached after the folding has taken place, see Fig.~\ref{fig:one}, left and right panels, respectively.

We call \emph{fold line} the planar curve $S \mapsto \bar{\bm{\gamma}}(S)$ where $S$ is the arc length.
We call \emph{ridge} or \emph{crease} its image after deployment by folding, which is a spatial curve $s \mapsto \bm{\gamma}(s)$, where $s$ is the arc-length.  Since the deployment map is a (continuous, piece-wise) isometry, then the map $s=s(S)$ is the identity and $s$ is arc-length on both $ \bar{\bm{\gamma}}$ and $\bm{\gamma}$.
The ridge $\bm{\gamma}$ separates two regions of a continuous surface with respective normals $\mb{N}^+$, $\mb{N}^-$, typically shaped as developable strips, that we call \emph{folds} or \emph{pleats}.

\begin{figure}
\begin{center}
%
%
\begin{tikzpicture}[scale=1]
\draw[thick] (8,1) rectangle (12,4);
\draw[very thick,->] (11,4) arc[start angle=90, end angle=270, radius=1.5] 
node[above]{$\bar\gamma(s)$};
\draw[->,very thick,blue] (9.5,2.5)--(10.5,3) node[right,above]{$\bar{\bm{r}}^+$};
\draw[->,very thick,red] (9.5,2.5)--(8.5,3) node[above]{$\bar{\bm{r}}^-$};
\draw[] (8.5,2.5)--(10.5,2.5);
\draw[dashed] (9.5,1)--(9.5,4);
\draw[red,<-] (9.,2.7) arc[start angle=150, end angle=280, radius=0.5]; 
\draw[red] (9,2.2) node[left]{$\bar\beta^-$};
\draw[blue,<-] (10,2.7) arc[start angle=30, end angle=-90, radius=0.5];
\draw[blue] (10,2.2) node[right]{$\bar\beta^+$};
\end{tikzpicture}
\hskip2cm
%
\begin{tikzpicture}[scale=1]
%
%
\filldraw[red]  (0.8,2) circle(.5pt);
\filldraw[red]  (5,4.5) circle(.5pt);
\draw[red,thick] (0.8,2)..controls(2,3.7)and(4,4.2)..(5,4.5);
\draw[red,thick] (1.8,1.44)--(0.8,2);
\draw[red,thick,dashed] (5.9,3.3)--(5.4,4);
\draw[red,thick] (5.4,4)--(5,4.5);
\draw[red,very thick,->] (1.8,1.44)--(2.9,0.7) node[below]{$\bm{u}^-$};
\draw[red,very thick,->] (1.8,1.44)--(2.45,2.6) node[above]{$\bm{N}^-$};

\filldraw[blue]  (2.9,2) circle(.5pt);
\filldraw[blue]  (6.4,4.2) circle(.5pt);
\draw[blue,thick] (2.9,2)..controls(3,2.4)and(4,3.9)..(6.4,4.2);
\draw[blue,thick] (5.9,3.3)--(6.4,4.2);
\draw[blue,->,very thick] (1.8,1.44)--(2.9,2) node[right]{$\bm{u}^+$};
\draw[blue,->,very thick] (1.8,1.44)--(1.3,2.5) node[above]{$\bm{N}^+$};

\filldraw  (1.8,1.44) circle(2.5pt) node[left=4pt,below]{$\bm{t}$};
\draw[thick] (1.8,1.44)..controls(3,3)and(4,3)..(5.9,3.3) node[below]{$\gamma(s)$};
\draw[->,very thick] (1.8,1.44)--(2.9,1.44) node[right]{$\bm{n}$};
\draw[->,very thick] (1.8,1.44)--(1.8,2.4) node[above]{$\bm{b}$};
\end{tikzpicture}
\end{center}
\caption{Local geometry of folding along a single curve. Left. Flat state before folding, with the 2D fold line $\bar\gamma$.
Right. Deformed state reached after the folding, with
the 3D crease $\gamma$.
Notation and terminology.
Frenet frame: $\bm{t}$, $\bm{n}$, $\bm{b}$.
Darboux frame: $\bm{t}$, $\bm{u}^\pm$, $\bm{N}^\pm$.}
\label{fig:one}
\end{figure}
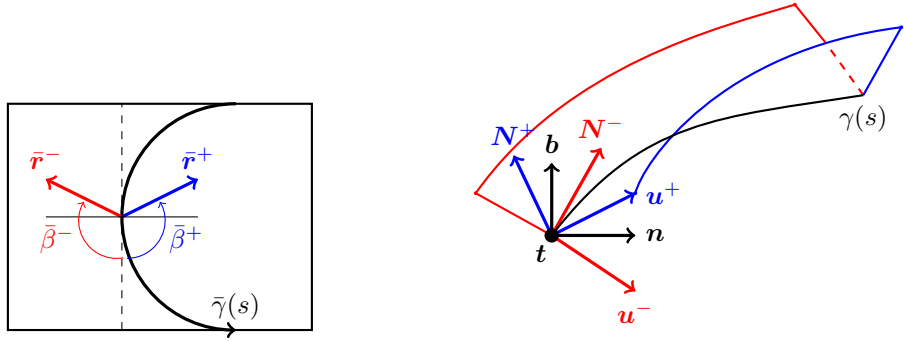

\begin{figure}
\begin{center}
\begin{tikzpicture}[scale=1]
\def\xa{0}; 
\def\xb{3};
\def\xc{7}; 
\def\xd{10};
%
%
%
\draw (\xa+1.5,3) node {Definition of $\alpha$};
\draw (\xc+1.5,3) node {Definition of $\beta$};
\draw[] (\xa,1) circle(3pt) node[left=6pt,below]{$\bm{t}$};
\draw[]  (\xa-1,1)--(\xa+1,1); \draw[]  (0,0.5)--(0,2);
\draw[very thick,->]  (\xa,1)--(\xa+1,1) node[below=4pt]{$\bm{n}$};
\draw[very thick,->]  (\xa,1)--(\xa,2)  node[above=2pt]{$\bm{b}$};
\draw[very thick,->,blue,rotate around={30:(0,1)}] (0,1)--(0,2) node[left]{$\bm{N}^+$};
\draw[very thick,->,blue,rotate around={30:(0,1)}] (0,1)--(1,1) node[right]{$\bm{u}^+$};

\draw[blue,thick,->] (0.7,1) arc[start angle=0, end angle=30, radius=0.7];
\draw (\xa,-0.5) node[blue]{$\alpha^+>0$};
\draw (\xb,1) circle(3pt) node[left=6pt,below]{$\bm{t}$};;
\draw (\xb-1,1)--(\xb+1,1);\draw (\xb,0.5)--(\xb,2);
\draw[very thick,->]  (\xb,1)--(\xb+1,1) node[above=4pt]{$\bm{n}$};
\draw[very thick,->]  (\xb,1)--(\xb,2)  node[above=2pt]{$\bm{b}$};
\draw[very thick,->,red,rotate around={-30:(\xb,1)}]
      (\xb,1)--(\xb,2) node[right]{$\bm{N}^-$};
\draw[very thick,->,red,rotate around={-30:(\xb,1)}] 
     (\xb,1)--(\xb+1,1) node[right]{$\bm{u}^-$};

\draw[red,thick,->] (\xb+0.7,1) arc[start angle=0, end angle=-30, radius=0.7];
\draw (\xb,-0.5) node[red]{$\alpha^-<0$};
\draw[red] (\xc,1) circle(3pt) node[right=10pt,above]{$\bm{N}^-$};;
\draw[]  (\xc-1,1)--(\xc+1,1);\draw[]  (\xc,0.5)--(\xc,2);
\draw[very thick,->]  (\xc,1)--(\xc+1,1) node[red, below=4pt]{$\bm{u}^-$};
\draw[very thick,->]  (\xc,1)--(\xc,0)  node[left=2pt]{$\bm{t}$};
\draw[very thick,->,red,rotate around={60:(\xc,1)}] (\xc,1)--(\xc,2) node[above]{$\bm{r}^-$};

\draw[red,<-] (\xc-0.5,1.3) arc[start angle=150, end angle=280, radius=0.5]; 
\draw (\xc,-0.5) node[red]{$\beta^-<0$};
\draw[blue] (\xd,1) circle(3pt) node[left=10pt,above]{$\bm{N}^+$};;
\draw[]  (\xd-1,1)--(\xd+1,1);\draw[] (\xd,0.5)--(\xd,2);
\draw[very thick,->] (\xd,1)--(\xd+1,1) node[blue,below=4pt]{$\bm{u}^+$};
\draw[very thick,->] (\xd,1)--(\xd,0)  node[left=2pt]{$\bm{t}$};
\draw[very thick,->,blue,rotate around={30:(\xd,1)}] (\xd,1)--(\xd+1,1) node[above]{$\bm{r}^+$};

\draw[blue,<-] (\xd+0.5,1.3) arc[start angle=30, end angle=-90, radius=0.5]; 
\draw (\xd,-0.5) node[blue]{$\beta^+>0$};
\end{tikzpicture} \end{center}
\caption{Definition of the angles $\alpha$ and $\beta$.}
\label{fig:two}
\end{figure}
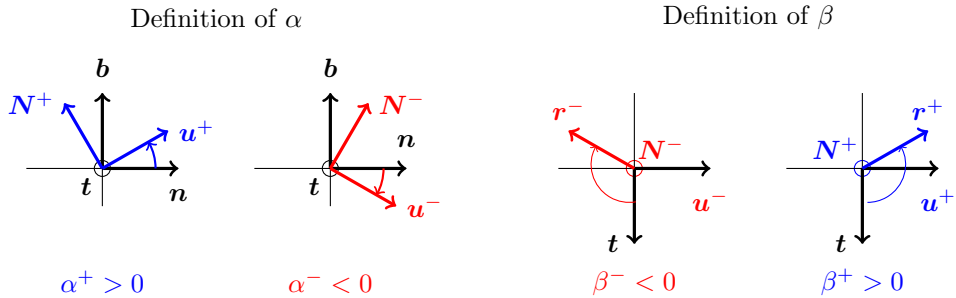

We denote the Frenet frame along $\bm{\gamma}$ with $\{ \mb{t}, \mb{n}, \mb{b}\} $ and the Darboux frames with
$\{ \mb{t}, \mb{u}^\pm, \mb{N}^\pm\} $. 
Let $\alpha^\pm$ be the angle that brings $\mb{n}$ onto $\mb{u}^\pm$, see Fig.~\ref{fig:two}; we have then
\begin{equation}
\mb{u}^\pm=\cos\alpha^\pm \mb{n} + \sin\alpha^\pm \mb{b}\,.
\end{equation}

The two surfaces joined by the ridge $\bm{\gamma}$ may be developable. In this case, we write them as
\begin{equation}\label{eq:pleat*}
\mb{x}(s,v)
= \bm{\gamma}(s) +v \mb{r}^\pm(s)\,,\quad v\in [0, \rm{width}^\pm >0 ]
\end{equation}
where the unit vector  $\mb{r}^+(s)$ (resp., $\mb{r}^-(s)$) gives the direction of the ruling at $\bm{\gamma}(s)$ on the positive (resp., negative) side of the surface. Denoting by $\beta^\pm(s)$ the angle that brings $\mb{t}(s)$ onto $\mb{r}^\pm(s)$,
the components of $\mb{r}^\pm(s)$ in the Darboux frames $\{ \mb{t}, \mb{u}^\pm, \mb{N}^\pm\} $ are $(\cos\beta^\pm, \sin\beta^\pm, 0)$
and hence
\begin{equation}\label{eq:rulings*}
\mb{r}^\pm(s)=\cos\beta^\pm \mb{t} + \sin \beta^\pm \left(\cos\alpha^\pm \mb{n} + \sin\alpha^\pm \mb{b} \right)\,.
\end{equation}
We recall the Frenet formulas for the ridge $\bm{\gamma}$ seen as a space curve 
\begin{eqnarray}\label{0002*}
\frac{d\bm{t}}{ds}= \kappa \bm{n}\,,\quad
\frac{d\bm{n}}{ds}= - \kappa \bm{t} + \tau \bm{b}\,, \quad
\frac{d\bm{b}}{ds}= - \tau \bm{n}
\end{eqnarray}
where $\kappa$ is the curvature and $\tau$ is the torsion, and the Darboux formulas for $\bm{\gamma}$ seen as a curve traced on either of the two surfaces with normal $\bm{N}^\pm$ (the pleats separated by $\bm{\gamma}$) 
\begin{eqnarray}\label{0004*}
\frac{d\bm{t}}{ds}= \kappa^\pm_g \bm{u^\pm} + \kappa^\pm_N \bm{N}^\pm \,,\quad
\frac{d\bm{u}^\pm}{ds}= - \kappa^\pm_g \bm{t} + \tau^\pm_r \bm{N}^\pm\,, \quad
\frac{d\bm{N}^\pm}{ds}= -\kappa^\pm_N \bm{t} - \tau^\pm_r \bm{u}^\pm
\end{eqnarray}
where $\kappa^\pm_g$, $\kappa^\pm_N$, and $\tau^\pm_r$  are geodesic curvature, normal curvature and  relative torsion of $\bm\gamma$ with respect to the surface with normal $\bm{N}^\pm$, see \cite{doCarmo76}. For the  planar curve $\bar{\bm{\gamma}}$ (the fold line, namely, the inverse image of the crease $\bm{\gamma}$  before folding)  formulas \eqref{0002*} and \eqref{0004*} 
become  
\be\label{0004*planar}
\frac{d\bar{\bm{t}}}{ds}= \bar{\kappa} \bar{\bm{u}} \,,\quad
\frac{d\bar{\bm{u}}}{ds}= - \bar{\kappa} \bar{\bm{t}} \,, \quad
\frac{d\bar{\bm{N}}}{ds}= 0
\ee
where $\bar{\bm{N}}$ is the normal to the plane containing the pleats before folding, and $\bar{\bm{u}}=\bar{\bm{n}}$ is the normal to $\bar{\bm{\gamma}}$.

Folding is realized by a continuous piece-wise isometry. This 
puts stringent constraints on the admissible folded geometries (kinematic compatibility conditions, akin to the Hadamard jump conditions on the gradients of a continuous piecewise affine map). Since isometries preserve distances, angles, and geodesic curvature, we have that
\begin{equation}
\beta^\pm=\bar{\beta}^\pm\,,\quad\kappa^\pm_g=\bar\kappa\,.
\end{equation}
Then, continuity of the folding map together with standard results of differential geometry imply 
\begin{equation}\label{0020*}
\kappa^-_N= \pm \kappa^+_N
\end{equation}
and 
\begin{equation}\label{0022*}
\alpha^-=-\alpha\,, \quad \text{where }  \alpha:=\alpha^+\,,
\end{equation}
so that the dihedral angle of the folded ridge, which we call the \emph{folding angle} $\theta$, is bisected by the normal $\mb{n}$ and can be written as
\begin{equation}\label{0024*}
\theta
:= \pi -(\alpha^+-\alpha^-)= \pi -2\alpha\,.
\end{equation}
We notice that $\alpha=\pm \pi/2$ corresponds to $\theta
=0, 2\pi$, respectively, and that
\begin{equation}\label{ineq*}
\kappa=\frac{\bar{\kappa}}{\cos\alpha}\ge\bar{\kappa}\,.
\end{equation}

Using \eqref{0022*},
and since $\tau_r/\kappa_N=\cot\beta$, 
 it follows that (see \cite{doCarmo76,fuchs_omnibus} for details):
$$
\cot \beta^\pm = \frac{\tau +(\alpha^\pm)^\prime}{\kappa^\pm_g \tan \alpha^\pm}= \pm \frac{\tau \pm (\alpha)^\prime}{\bar{\kappa}\tan \alpha}
$$
and, by adding and subtracting, we deduce that
\begin{equation}\label{0030*}
\cot\beta^+ + \cot\beta^-= \frac{2}{\bar{\kappa}} \frac{\alpha^\prime}{\tan\alpha}\,,
\end{equation}
\begin{equation}\label{0031*}
\cot\beta^+ -  \cot\beta^-= \frac{2}{\bar{\kappa}} \frac{\tau}{\tan\alpha}\,.
\end{equation}

\begin{figure}[t]
\center
\includegraphics[width=0.4\columnwidth]
{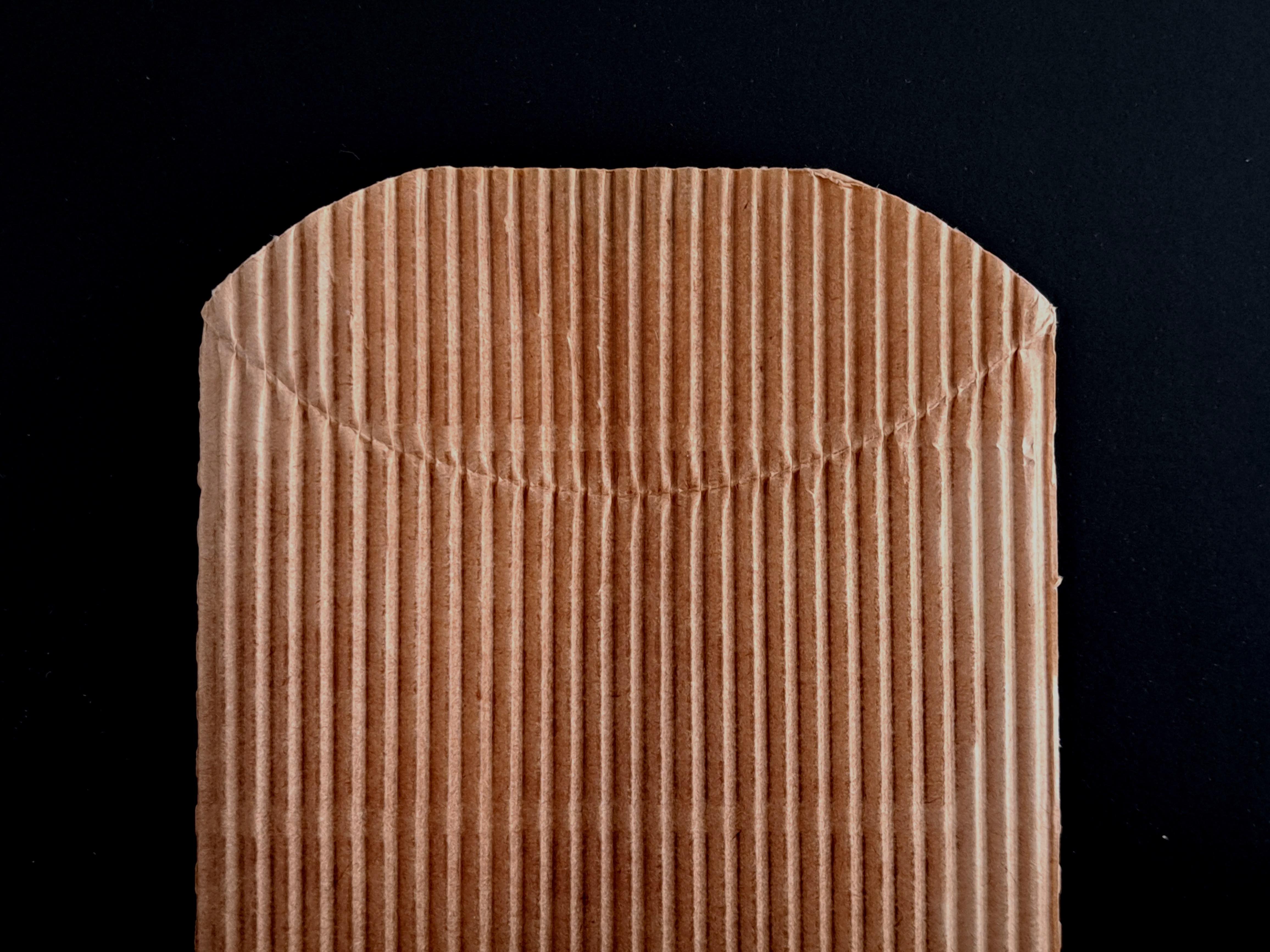}
\hskip1cm
\includegraphics[width=0.4\columnwidth]
{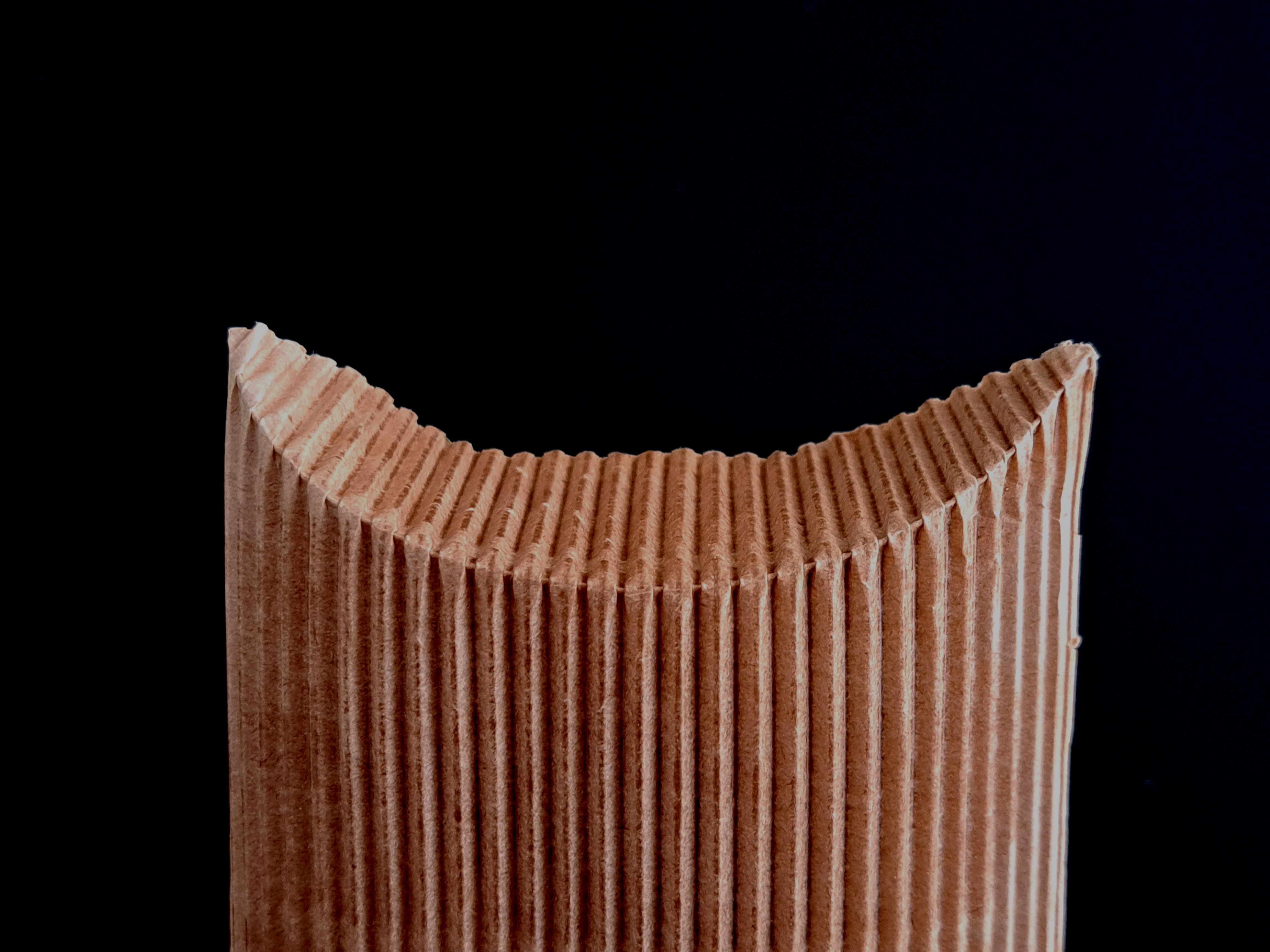}
\caption{Example of circular arc fold line with corrugated cardboard. Flat configuration (left) and
curved one (right).}
  \label{fig:figure002}
\end{figure}

Consider now the special case of a piecewise isometry in which the rulings are collinear in the flat state ($\bar{\beta}^+ - \bar{\beta}^-=\pi$). An example of a physical system realizing this condition is given by foldable containers made of corrugated cardboard, {\bf{see Figure~\ref{fig:figure002}}}.  Since angles on a surface are preserved by isometries, we also have $\beta^\pm=\bar{\beta}^\pm$ and it follows from \eqref{0030*},\eqref{0031*}
that $\tau=0$ (hence the folded ridge is always a planar curve!) and
\begin{equation}\label{0130*}
\frac{\alpha^\prime}{\tan\alpha}=\frac{\bar{\kappa}}{\tan\beta}=  \frac{|\beta^\prime|}{\tan\beta}\,,
\end{equation}
where $\beta:=\bar{\beta}^+$. ODE \eqref{0130*} is easily integrated and the profile $\alpha(s)$ is uniquely determined by the initial condition $\alpha(0)=\alpha_0$.
This single scalar parameter, that controls the folding angle $\theta=\pi-2\alpha$ at one point ($s=0$) of the ridge, uniquely determines the curvature $\kappa$ of the planar ridge $\gamma$ via \eqref{ineq*} ($\kappa=\bar{\kappa}/\cos\alpha$). The curve  $\gamma$ is then recovered (modulo rigid motions) from the knowledge of $\kappa$ and $\tau=0$.
The whole pleated structure (one fold) is then reconstructed from \eqref{eq:pleat*}. In this way a one-parameter family of curved creases is obtained, parameterized by the angle $\alpha_0$, which can be continuously unfolded to the flat state (that is, a one-DoF mechanism).
\begin{figure}[t]
\center
\includegraphics[width=0.9\columnwidth]{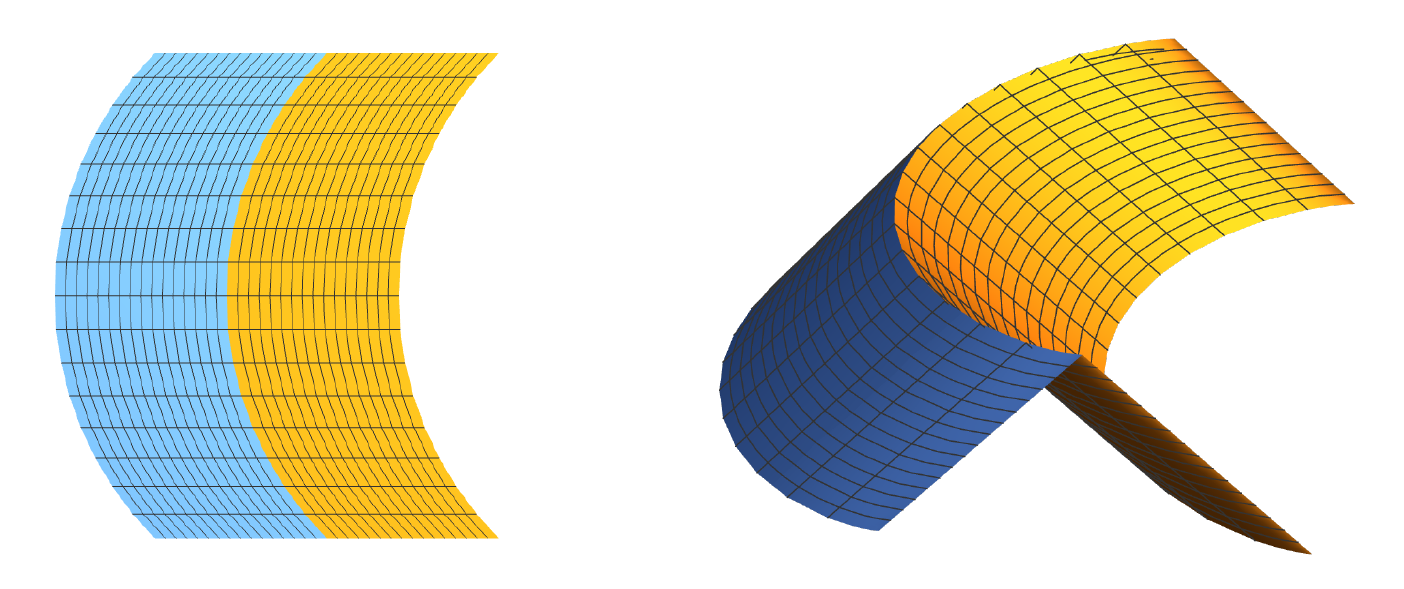}
\caption{Example of circular arc fold line and folded crease obtained with the geometric approach.}
  \label{fig:figure001}
\end{figure}

The existence of solutions of ODE \eqref{0130*} in a neighborhood of $s=0$ guarantees that the sheet can be folded locally along the prescribed fold line $\bar{\bm{\gamma}}$. The next question is the maximal interval of existence over which this solution can be continued, i.e. the maximal arc length distance $s^*$ over which the sheet can be folded along $\bar{\bm{\gamma}}$.
To explore this question in a concrete example, we consider the special case of a fold line shaped as a circle arc, set $\beta(0)=\beta_0=\pi/2$ (so that
$\beta^\prime<0$ in a neighborhood of $s=0$ and the fold line is symmetric with respect to the normal $\bar{\bf{n}}(0)$, which leads to a symmetric ridge with respect to the plane generated by ${\bf{n}}(0)$ and {\bf{b}}(0)), and we impose a folding angle $\theta=\pi/2$ at $s=0$ (hence $\alpha_0=\pi/4$). By integrating \eqref{0130*} in $(0,s)$ we obtain 
\begin{equation}\label{eq:pret_a_manger*}
\sin\alpha(s)=\sin\alpha_0\frac{1}{\sin\beta(s)} \quad \text{  (for  } \beta_0=\frac{\pi}{2})
\end{equation}
where $\beta(s)=\pi/2-s/R$, and which admits solution with real $\alpha(s)$ for any $\alpha_0\in (-\pi/2,\pi/2)$ as long as 
$\sin\beta(s)\ge |\sin\alpha_0|$.
At $s=s^\star$ where equality holds, $\alpha(s^\star)=\pm \pi/2$ and the folding angle of the ridge either vanishes or is equal to $2\pi$. We write
\begin{equation*}
\alpha(s)=\arcsin \left( \frac{\sin\alpha_0}{\sin\beta(s)}\right)\,,\quad s\in (0,s^*)
\end{equation*}
and the solution above cannot be continued past $s=s^\star$ without violating non-interpenetration of the two pleats joined by the ridge. For the case $\beta_0=\pi/2$, this occurs with $\beta(s^\star)=|\alpha_0|$. 
{\bf{Figure~\ref{fig:figure001}}}
shows a fold line $\bar{\bm{\gamma}}$ shaped as a circular sector (arc of circle) of radius $R$. If $\alpha_0$ has to be varied in the interval $(0,\pi/4)$, so that the mechanism at $s=0$ works as a hinge with folding angle $\theta$ closing from $\pi$ (flat) to $\pi/2$ (in which case the two sides of the ridge are `orthogonal' at $s=0$), then the maximal length of the fold-line is $s^*=R \pi/4$ corresponding to a circular sector with  opening angle $\pi/4$ (or $\pi/2$ for its symmetric version).

In practice, the geometric construction described above can be 
realized in the folding of a sheet if the isometry is rigidly enforced (i.e., if the sheet is unstretchable), and if the direction of the rulings can be prescribed a-priori. This second condition can be guaranteed if the bending energy of the sheet is anisotropic, 
so that the sheet is inflexible in one direction (the one of the rulings \eqref{eq:rulings*}) and flexible in  the perpendicular one with either zero or finite bending modulus $B\geq 0$. This can be obtained, for example,  with corrugated cardboard ({\bf{Fig.~~\ref{fig:figure002}}}), with sheets containing embedded stiff and aligned fibers or, to some extent, with sheets made of inflated parallel and unstretchable tubes (see \cite{gao2020shape}).

%
\subsection{Mechanical modeling}\label{sec:elastic}
%
We consider two mechanical models: 
1) 2D shell model; 2) 3D solid mechanics.
The two models share the same geometric footprint, that is a flat
rectangular surface $\mathcal{S}=L\times W$, with the length $L$ and
the width $W$ parallel to the $X$ and $Y$ directions, respectively. 
The 2D region $\mathcal{S}$ is first split into pleats by fold lines
(one fold is considered in this section, two and three in the following) and then extruded along the $Z$ direction
to a 3D region of thickness $H$ defined by
$\mathcal{V}=\mathcal{S}\times (-H/2,+H/2)$.
This 3D region $\mathcal{V}$ represents the reference configuration
of both the 3D solid and the shell-like solid; see
Sections~\ref{sec:3D_model} and \ref{sec:shell_model} for a brief summary of the theory, and
Section~\ref{sec:geometries} for the details about the geometries.

Mechanical coupling between folds is enforced through kinematic constraints. Adjacent folds are constrained to exhibit identical displacements along the fold line joining them. To enforce this constraint numerically, we observe that each physical fold line is a boundary curve for two adjacent folds, which we conventionally call the left and right folds and fold lines. 
Denoting the displacements of the left and right folds as $\mb{u}_L$ and $\mb{u}_R$, respectively, 
we impose the constraint $\mb{u}_R=\mb{u}_L$ on the right fold line;
reaction forces are then applied to the left fold line to maintain this coupling.

\begin{figure}[t]
\center
\includegraphics[width=0.4\columnwidth]{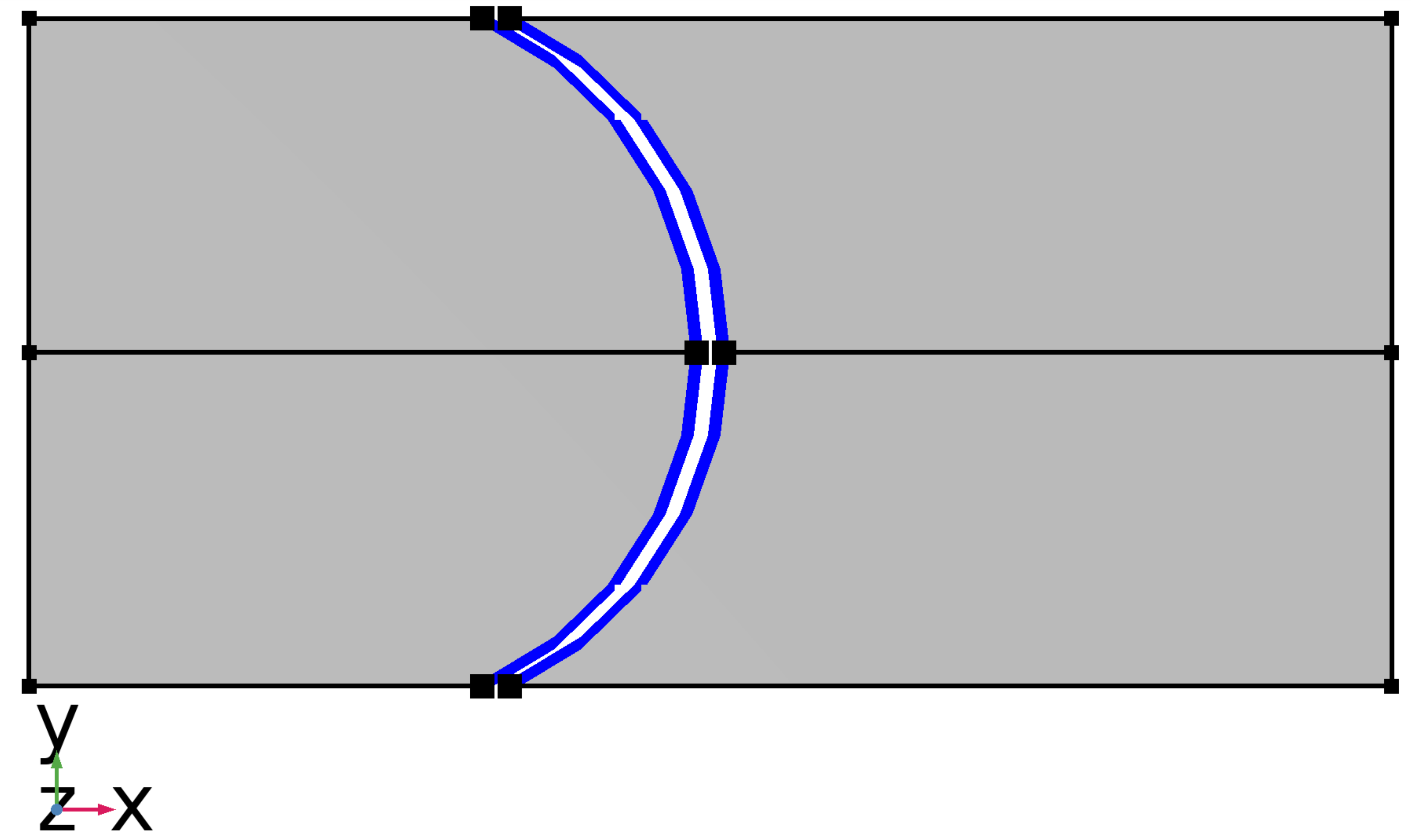}
\includegraphics[width=0.5\columnwidth]{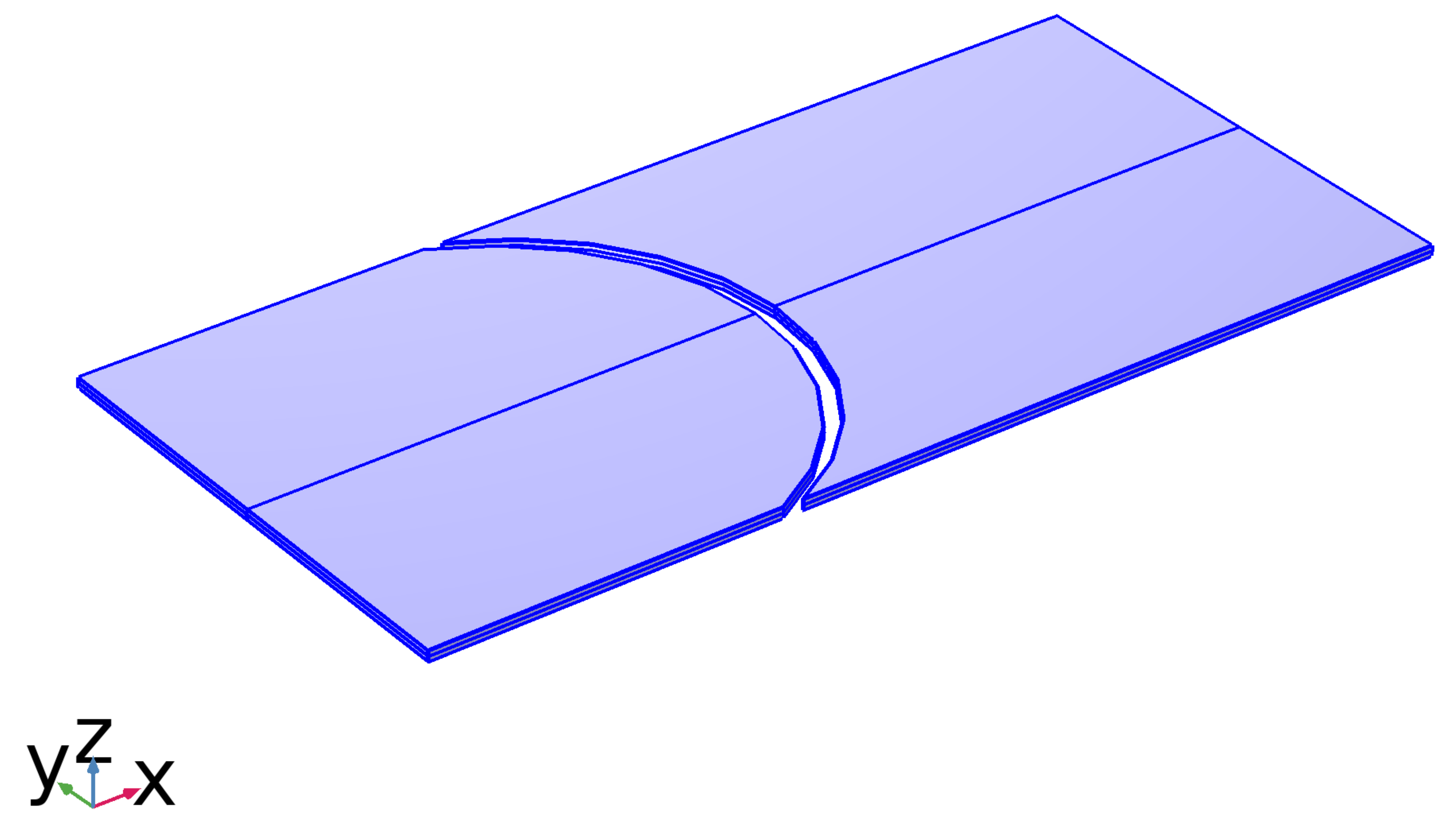}
\caption{Geometries with two folds. 
Left: 2D geometry with two separate folds; the mechanical coupling is obtained
by kinematical constraints on the two adjacent fold-lines (blue).
Right: 3D geometry extruded from the 2D one. The mechanical coupling is obtained
by the same kinematical constraints on the fold-lines used for the 2D case.}
  \label{fig:1F_geometries}
\end{figure}

It is worth mentioning that a key characteristic of the constraint is that it solely restricts relative translational motion between the two folds, allowing for independent rotational movement.
This is because the shell model's rotational state variables are unconstrained, and controlling linear displacement along a line in the solid model does not restrict relative rotations.
Figure~\ref{fig:1F_geometries} shows the case with two folds; on the left we have the 2D geometry, with  fold lines highlighted in blue, and on the right the 3D one.

The next two subsections provide detailed insight into the specific problems
we addressed to compare the two modeling approaches. To assess the effectiveness of the models, we conducted tests on various geometries featuring two folds and a circular fold line with varying curvature radii, $R_o$.
We always use the same flat region $\mathcal{S}=L\times W$, with $L=2\,W$, and thickness $H=L/100$,
using a different radius of curvature of the circular fold line: $R_o=0.6\,W,\ldots,1.1\,W$.

\begin{figure}[t]
\center
\includegraphics[width=0.3\columnwidth]{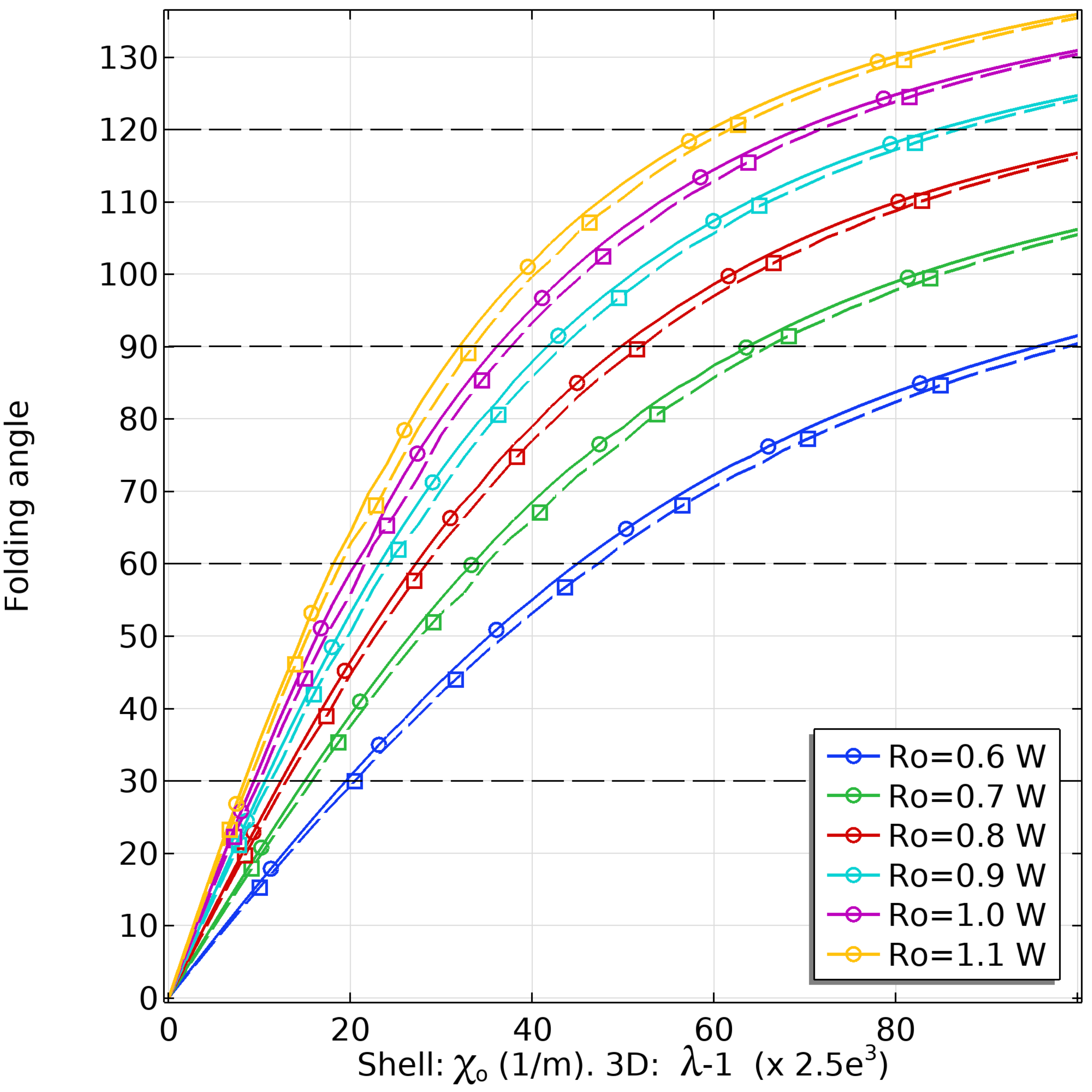}
\includegraphics[width=0.3\columnwidth]{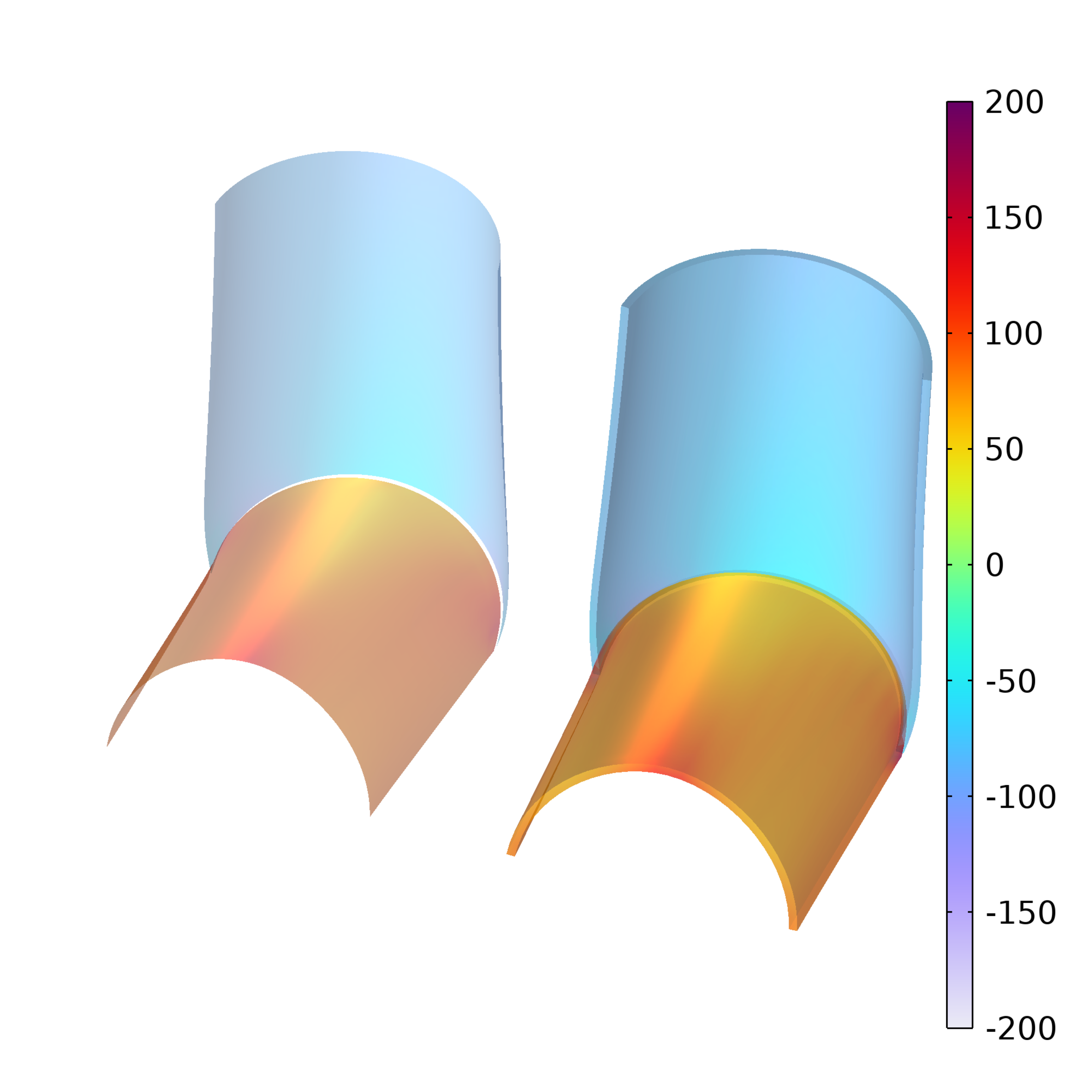}
\includegraphics[width=0.3\columnwidth]{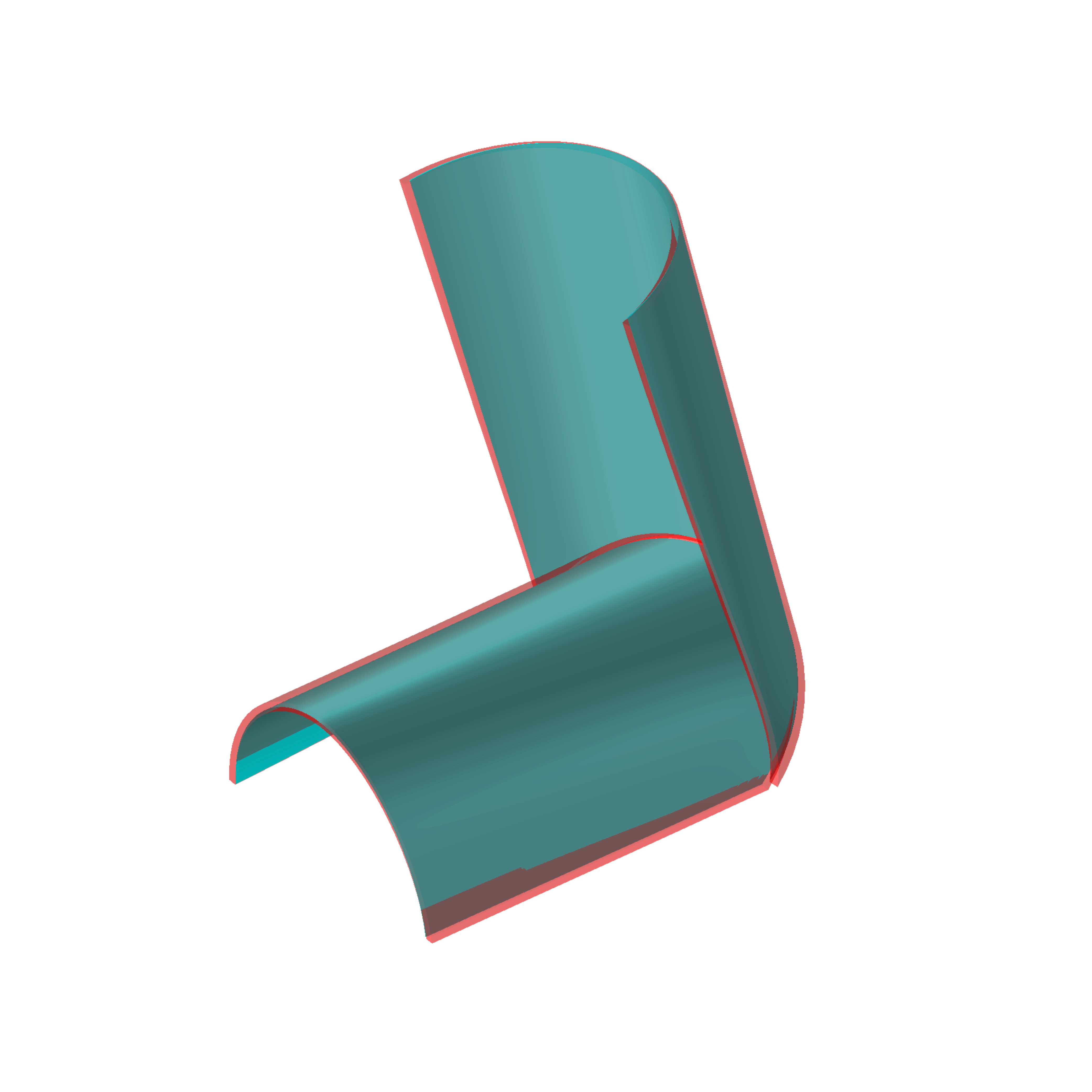}
\caption{Comparison of shell model with 3D solid model.
Left: Folding angle $\theta_{12}$ versus the target curvature $\chi_o$ for the shell model (solid, circle),
and versus the target stretch $\lambda_\|-1$ for the 3D model (dashed, square). To align the 
x-coordinates we rescale the stretch by $2.5\,10^3\,(\lambda_\|-1)$.
Center: two configurations for $R_o=0.7\,W$, at $\chi_o=100$ 1/m and $\lambda_\|-1$=0.04, 
side by side (left=shell; right=3D).
Right: the previous two configurations superimposed each other.}
  \label{fig:1F_geometries_2}
\end{figure}

\begin{figure}[t]
\center
\includegraphics[height=0.22\columnwidth]{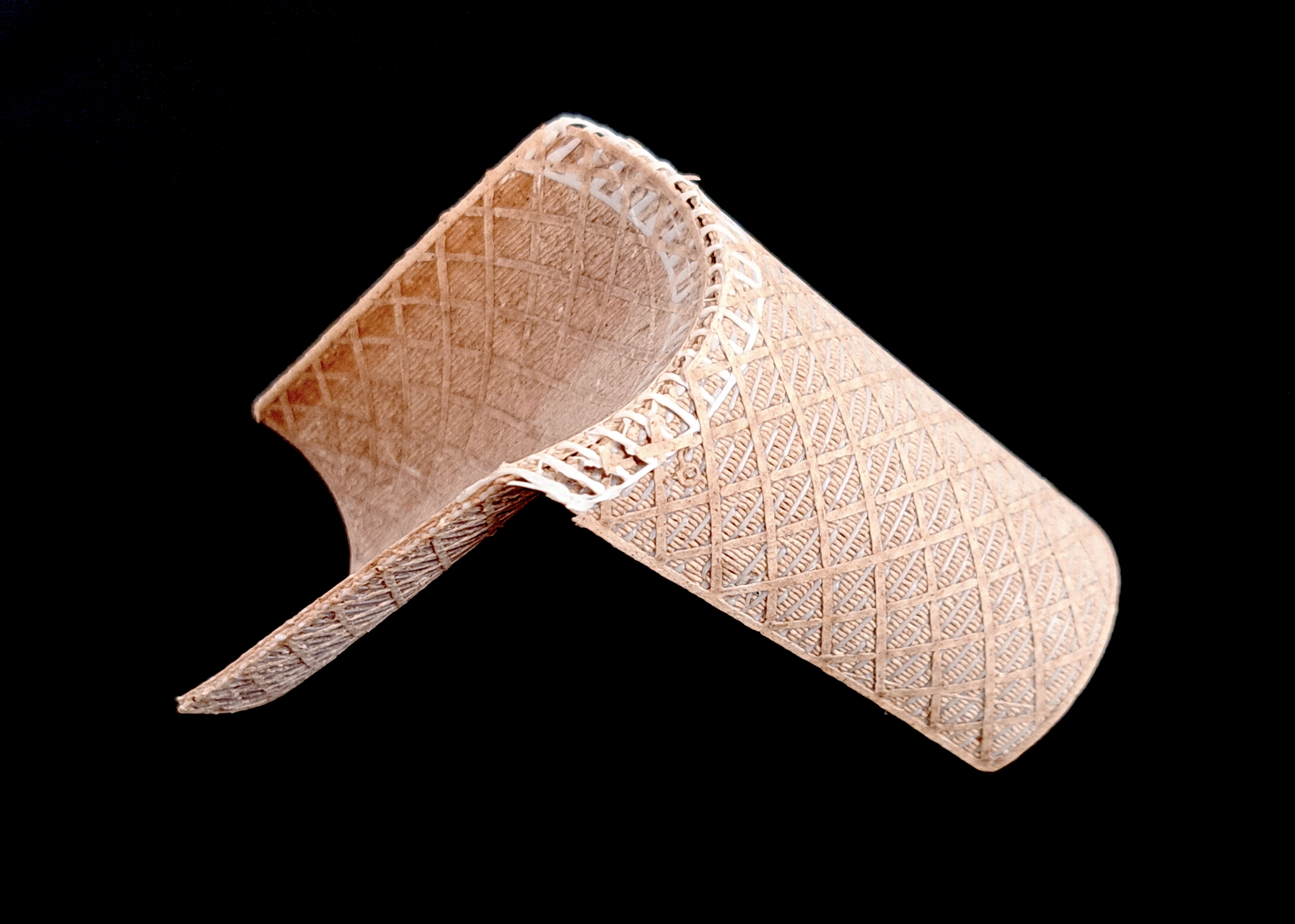}
\hskip0.1cm
\includegraphics[height=0.22\columnwidth]{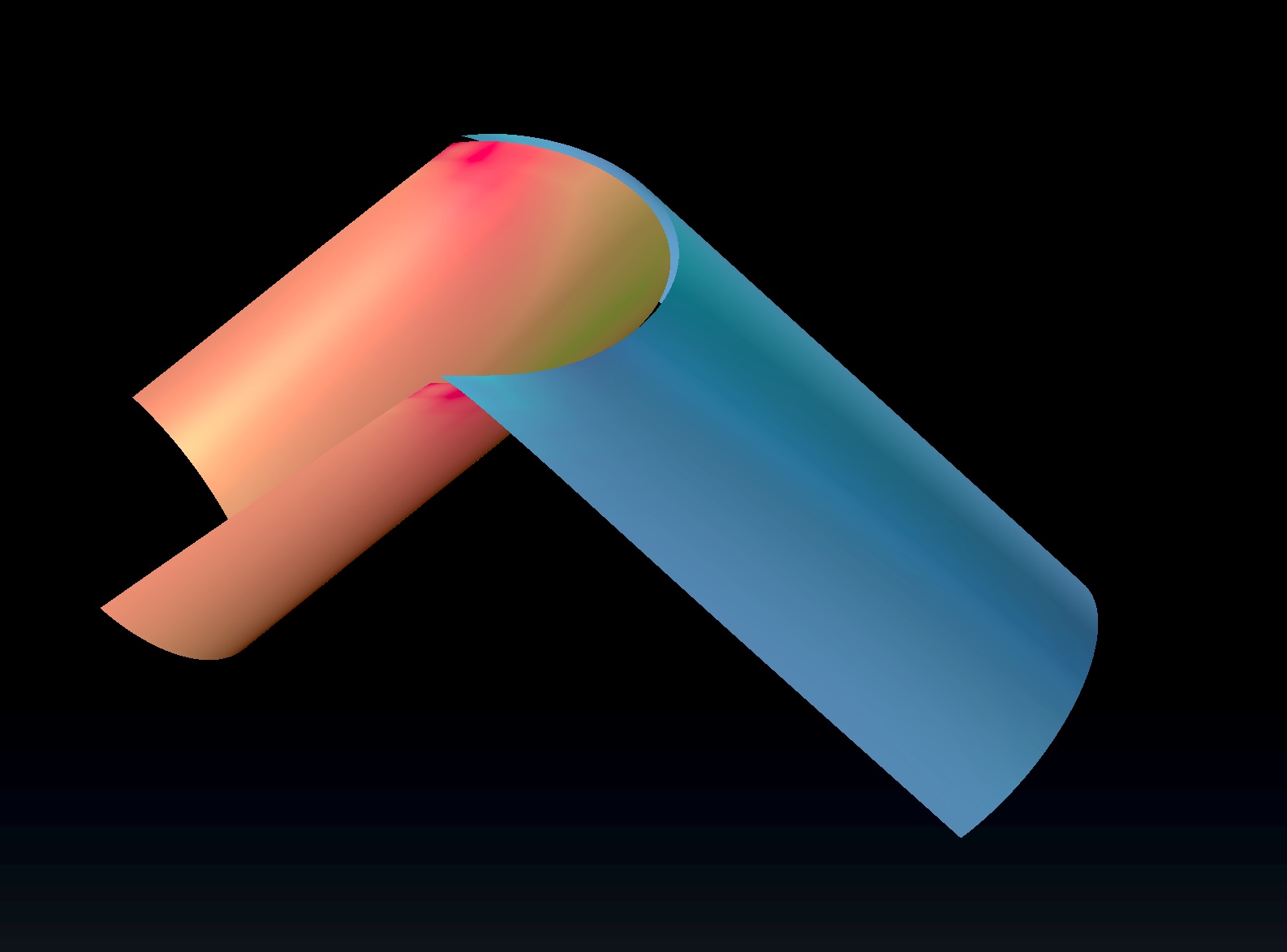}
\hskip0.1cm
\includegraphics[height=0.22\columnwidth]{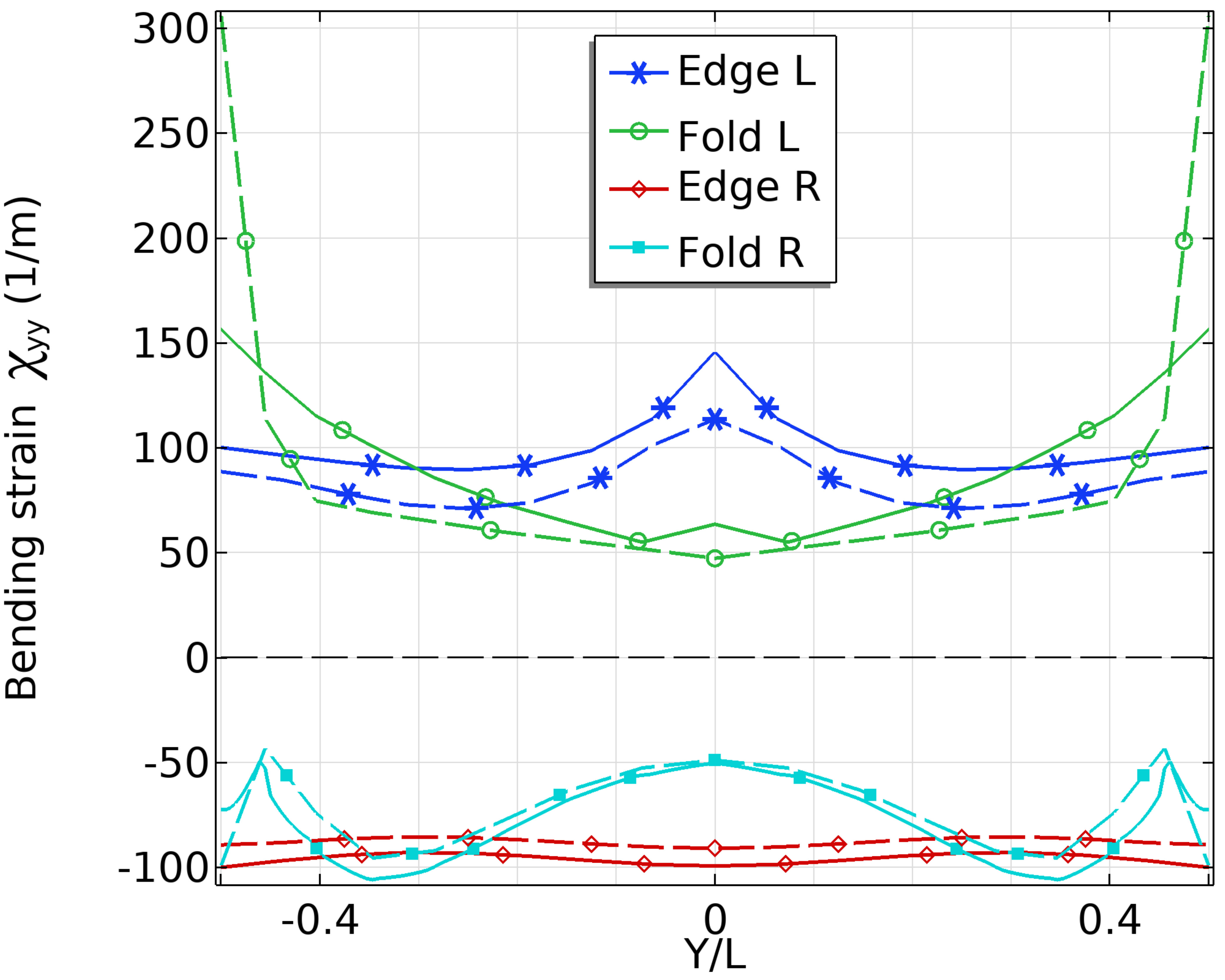}
\caption{Left. The hygromorph, courtesy of T. Cheng and Y. Tahouni. Center. Result from shell model; $R_o=0.6\,W$, $\chi_o=100$ 1/m. Right. Bending strain $\chi_{yy}$ versus $Y/W$ along four lines; left (red fold) and right (blue fold) refers to the image at center. The solid lines correspond to the results obtained from the shell model, whereas the dashed lines correspond to the results from the 3D solid model.}
  \label{fig:figure003}
\end{figure}

The specific test geometry considered in these subsections allows us to address the origin of the geometric  `tapering' predicted by both  mechanics models and observed experimentally in synthetic hygromorphic structures, see {\bf{Figure~\ref{fig:figure003}}}. The prescribed (constant) target curvature is achieved at the edges of the rectangular cross section farthest away from the fold line, see right panel of Fig.~\ref{fig:figure003}.
Moving towards the fold line, the observed curvature decreases to a reduced value, due to the interference of the fold line, which favors different (nonconstant) values of curvature, as shown by the geometric approach in Section \ref{sec:geometric_approach}. The existence of fold lines that do not interfere with the prescribed target curvature is further investigated in Section~\ref{sec:circ_vs_trig}.

\subsubsection{Approach by active shell models: actuation by target curvature}\label{sec:shell}
We solve the weak equation (\ref{shell_weak}), using as only input the target curvature
$\chi_o$; the deformed configurations are obtained by assigning the
target curvature $\chi_o$ with only one non trivial component; 
we have:
\be\label{kappa_o}
\chi_o=\left[\begin{array}{cc} 0 &  0 \\[2mm]
                               0 & \chi_{o,yy} 
    \end{array}\right];
\quad\textrm{thus, }
\chi_e=\left[\begin{array}{cc} 
\chi_{xx} &  \chi_{xy} \\[2mm]
\chi_{xy} & \chi_{yy}-\chi_{o,yy} 
    \end{array}\right].
\ee
The bending term in the shell energy \eqref{shell_energy} is zero when $\chi_e=0$, that is when the actual curvature is equal to the target one; in general, this equality cannot be realized, and the minimum of the energy is obtained with all the three terms different from zero; in our models, the bending
energy is the leading one.

In particular, for each shell, we solve for the configurations $\mathcal{C}_i$
corresponding to a sequence of $n$ pairs $K_{i}=(\chi_{oi},-\chi_{oi})$,
where $\chi_{oi}$ are the target curvatures assigned to the first fold
and  $-\chi_{oi}$ the ones of the second fold;
$\chi_{oi}$ ranges in $(0, \chi_{m})$ (1/m).

\subsubsection{Approach by 3D active elasticity models: actuation by target metric}
We solve the weak equation (\ref{3D_weak}), using as only input a target strain. 
The deformed configurations are obtained by assigning a 
target strain $\mb{E}_o$ parametrized by the stretch $\lambda_\|$, see (\ref{lambda_o}).
In particular, each of the two folds is split by a horizontal plane of symmetry in the bottom
and top halves; we set $\lambda_\|=1$ in the bottom-left and top-right regions, and we 
solve for the configurations $\mathcal{C}_i$
corresponding to a sequence of $n$ target stretches $\lambda_{i}=\lambda_{\| i}$,
with $\lambda_{1}=1$ and $\lambda_{n}=\lambda_{max}>1$; we use a minimal set of kinematical constraints that rule out rigid motions without inducing reactive forces.
%

\section{Role of geometry of a single fold-line: circular arcs vs trigonometric functions}
\label{sec:circ_vs_trig}
Following the modeling approach described
in Sec.~\ref{sec:shell}, 
we compare the behavior of flat rectangular shells $\mathcal{S}=L\times W$ 
split in two folds by different fold lines:
1) circular arc; 2) trigonometric function.
The geometries are described in Sec.~\ref{sec:geometries}.
The shells are not loaded, and have compatible constraints, that is
the deformations they exhibit do not generate reaction forces; the only input is the 
target curvature $\chi_o$, see
(\ref{kappa_o}).

For both shells, we solve for the configurations $\mathcal{C}_i$
corresponding to a sequence of $n$ pairs $X_{oi}=(\chi_{oi},-\chi_{oi})$,
where $\chi_{oi}$ are the target curvatures assigned to the first fold
and  $-\chi_{oi}$ the ones of the second fold;
$\chi_{oi}$ ranges in $(0, \chi_{m})$ (1/m).

We discuss the results showing: 1) the folding angle 
$\theta_{12}=2\,\alpha_{12}$ between the first and the second fold.
2) the actual curvature $\chi_{yy}$ of the deformed configuration $\mathcal{C}$
evaluated along the center line;
3) the elastic energy density (energy per unit area).

\begin{figure}
\center
\includegraphics[width=0.3\columnwidth]{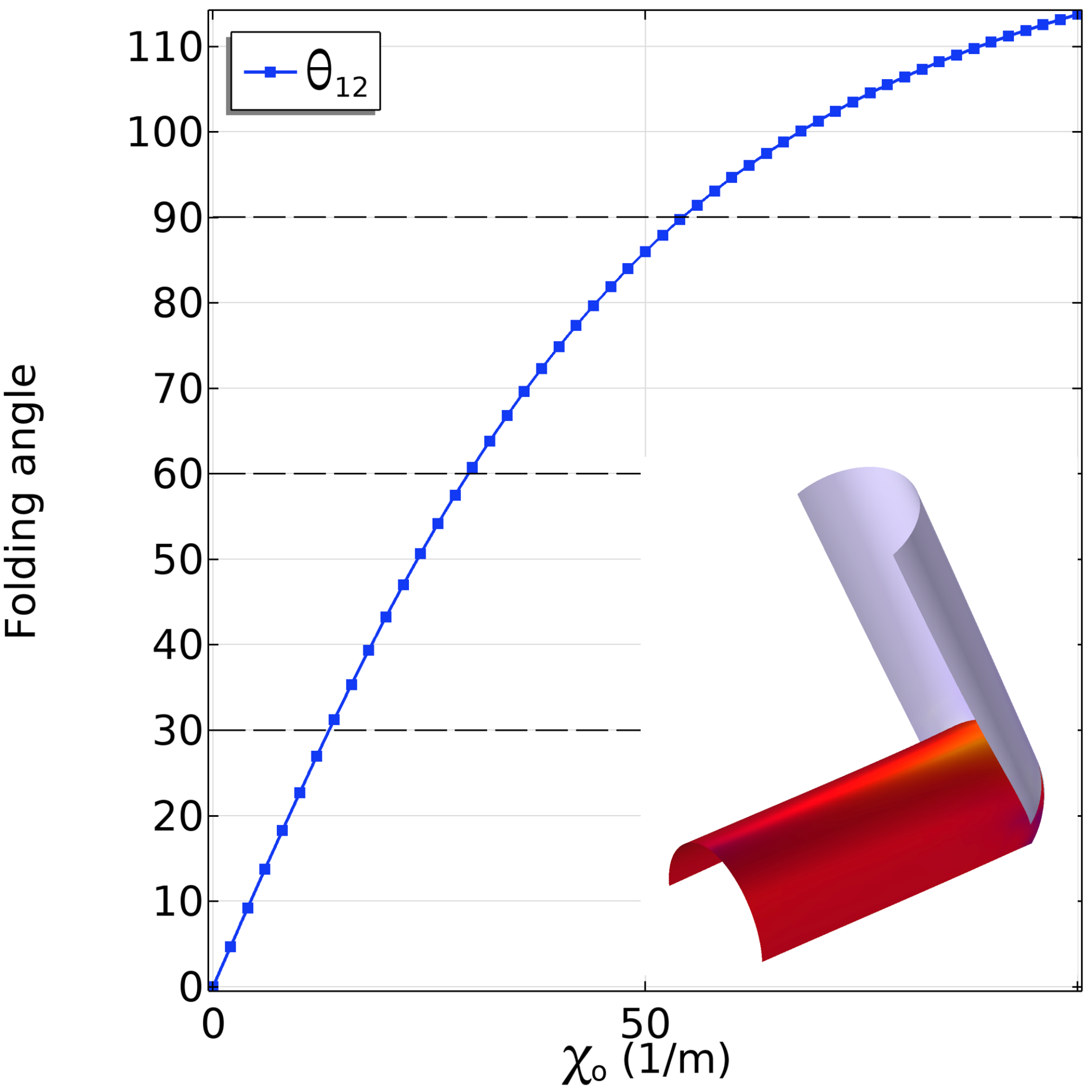}
\includegraphics[width=0.3\columnwidth]{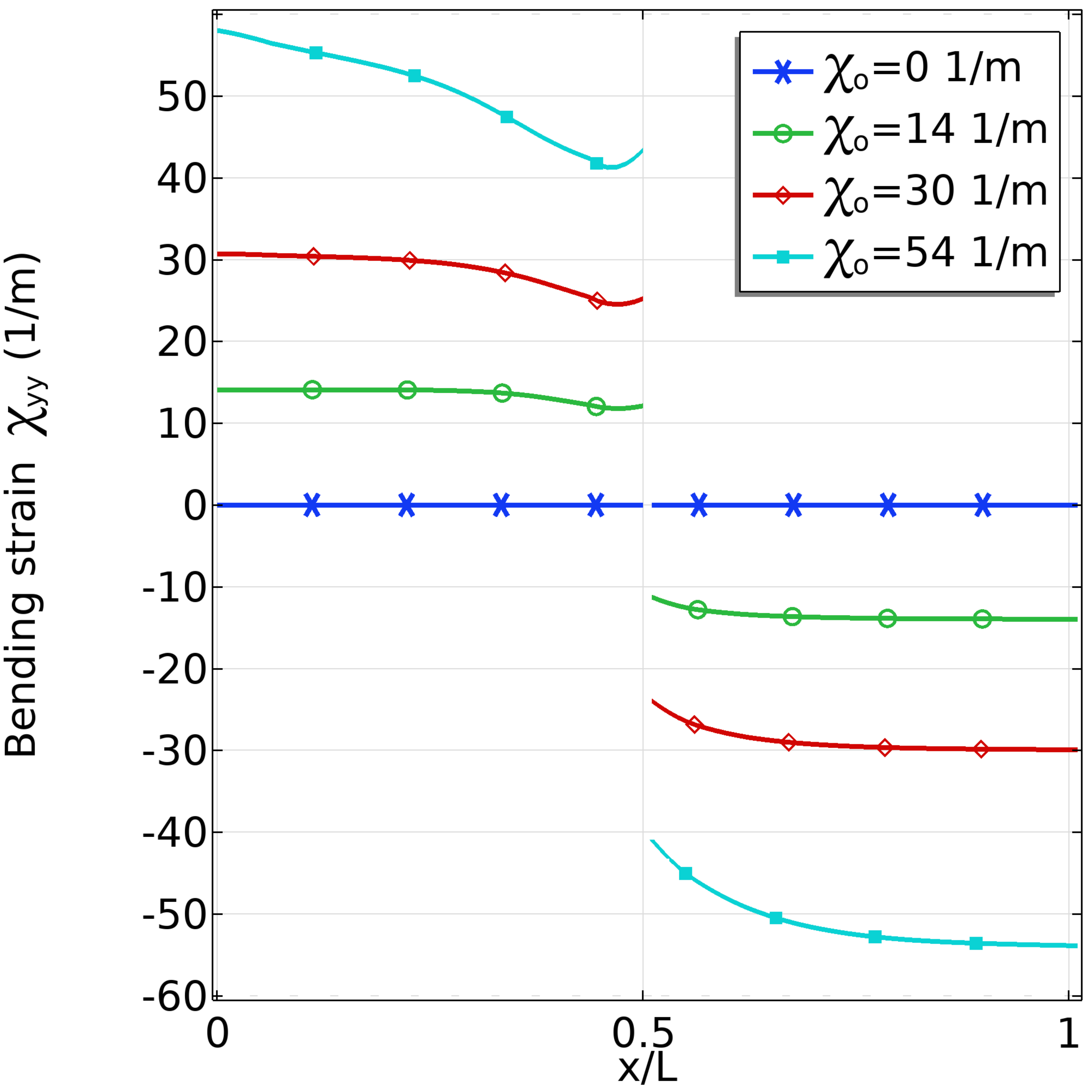}
\includegraphics[width=0.3\columnwidth]{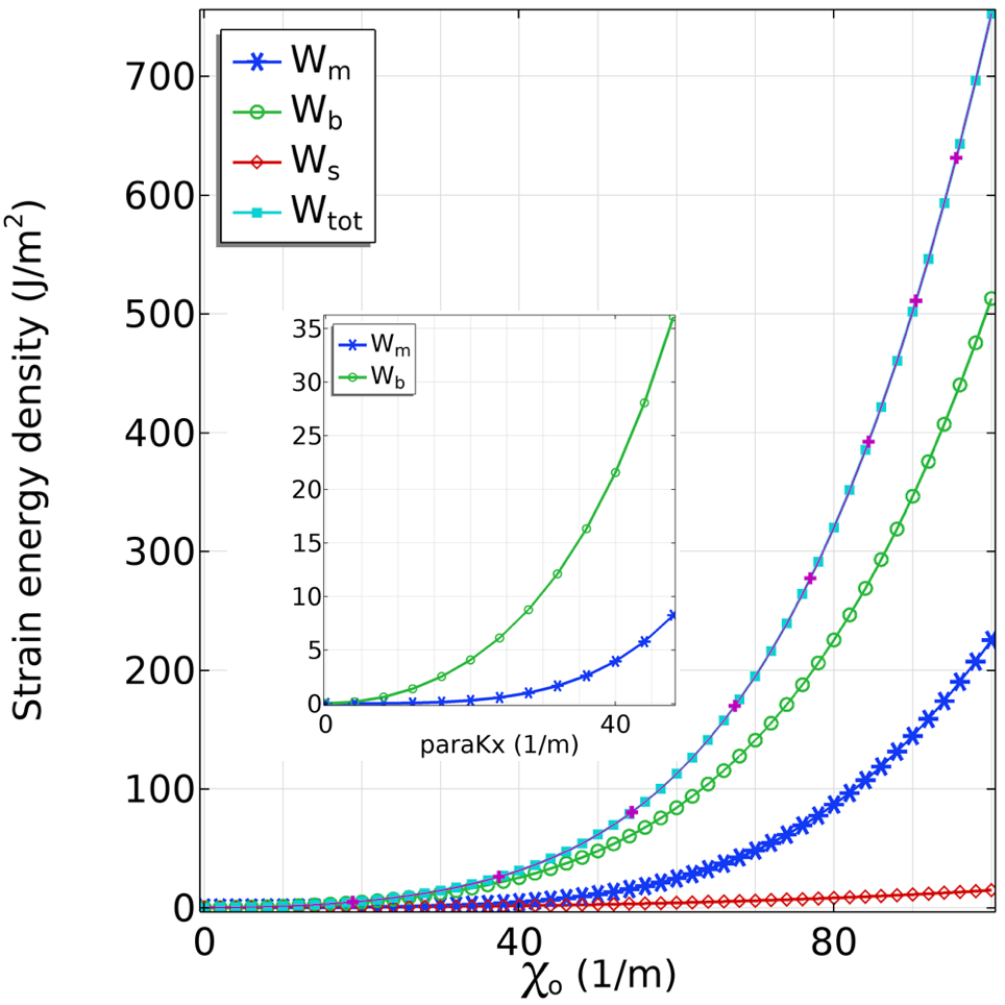}
\caption{Result for circular fold-line. 
Left: Folding angle $\theta_{12}$ between the two folds; the inset shows the configuration at $\chi_o=100$ 1/m. Center: Current bending strain 
$\chi_{yy}$ versus $X/L$ along the center line of the shell for three different values of the target curvature $\chi_o$ evaluated at
the folding angles $\theta_{12}=30^o, 60^o, 90^o$. 
It might be noticed that current $\chi_{yy}$ differs from 
the target curvature $\chi_o$ as consequence of the incompatibility of the curvature assigned to the two folds. The difference between $\chi_{yy}$
and $\chi_{o}$ increases with the folding angle.
Right: Elastic energies
versus target curvature $\chi_o$; 
energies increases with $\chi_o$, a consequence of the aforementioned
incompatibility of $\chi_o$.}
  \label{fig:1F_circ_theta_chi_ene}
\end{figure}

\begin{figure}
\center
\includegraphics[width=0.3\columnwidth]{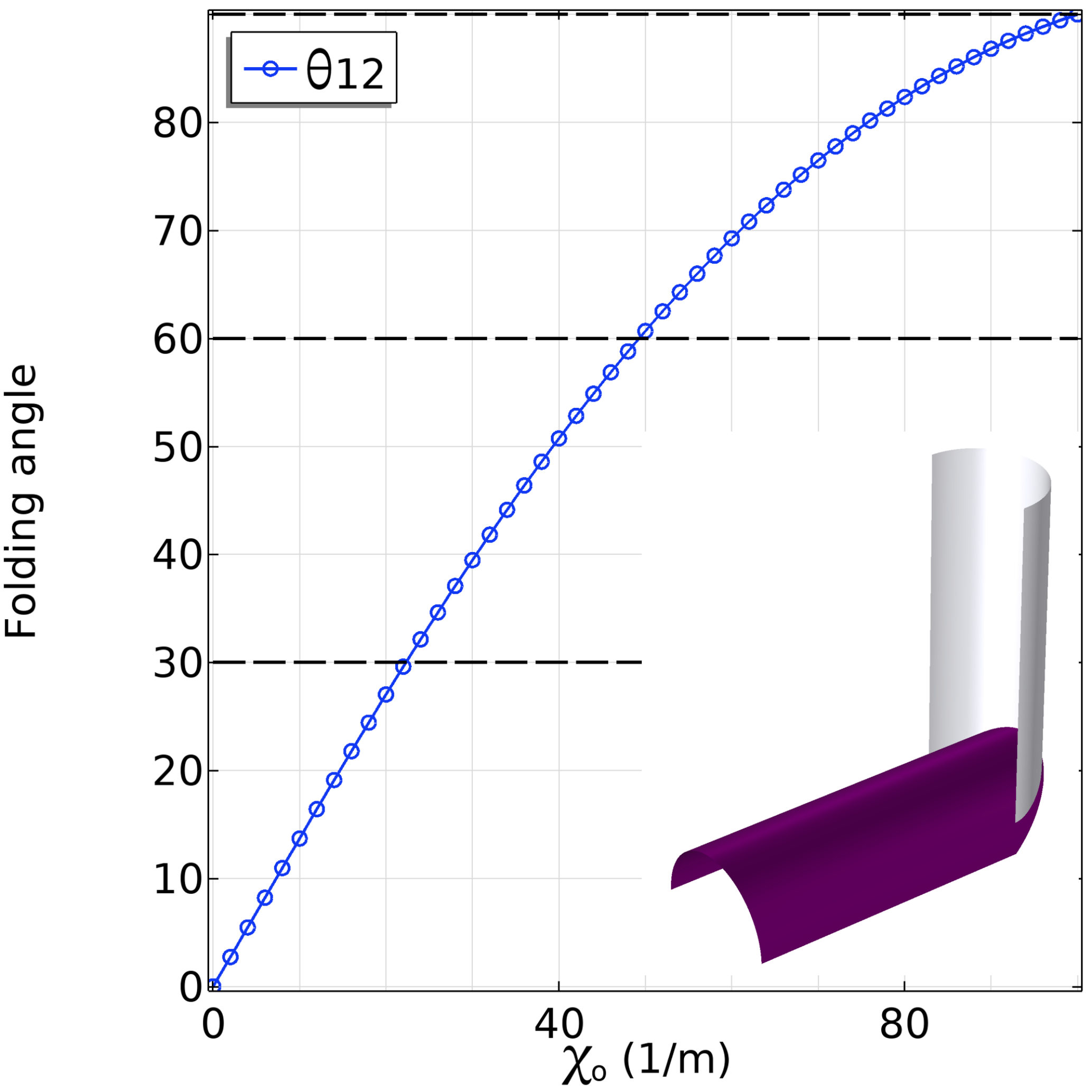}
\includegraphics[width=0.3\columnwidth]{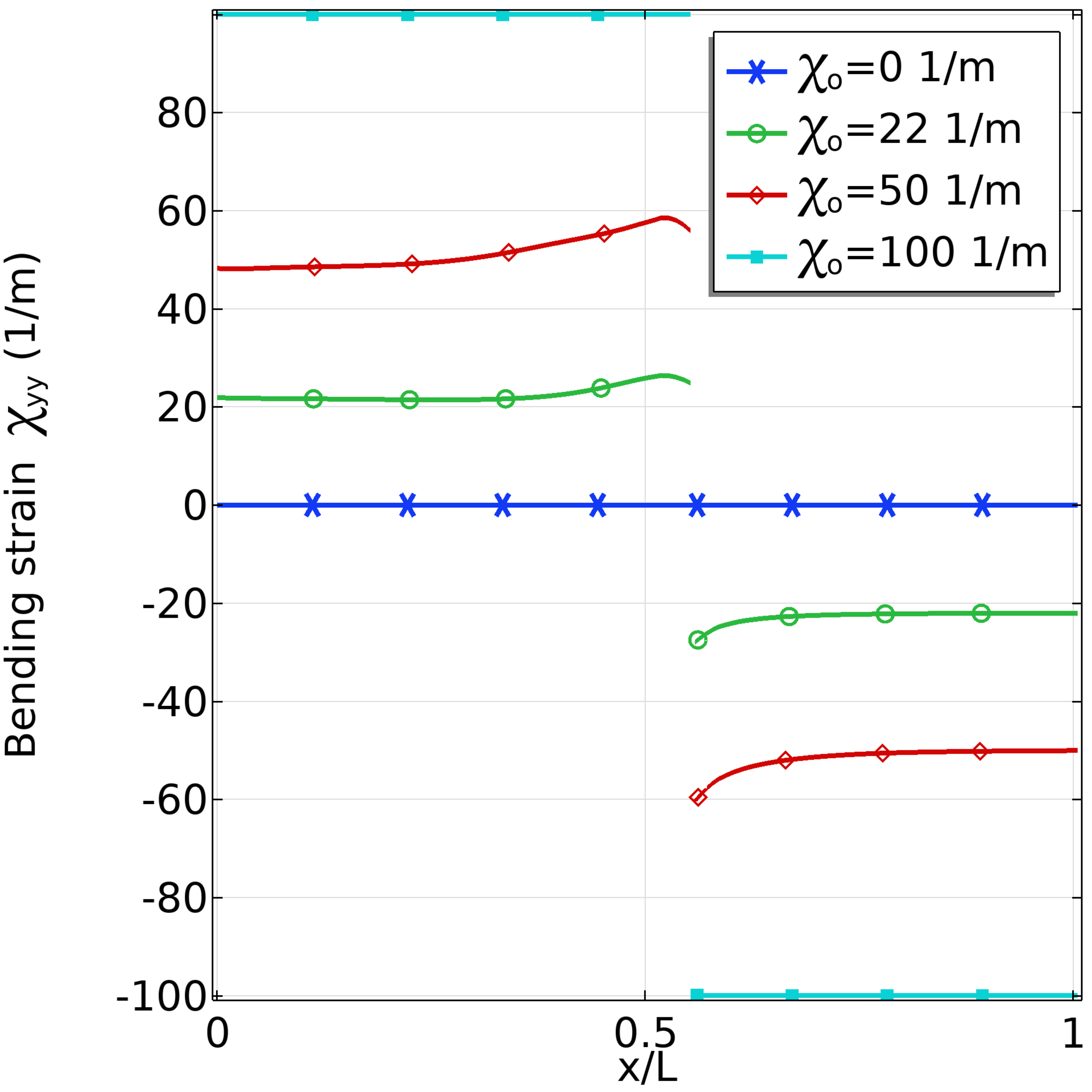}
\includegraphics[width=0.3\columnwidth]{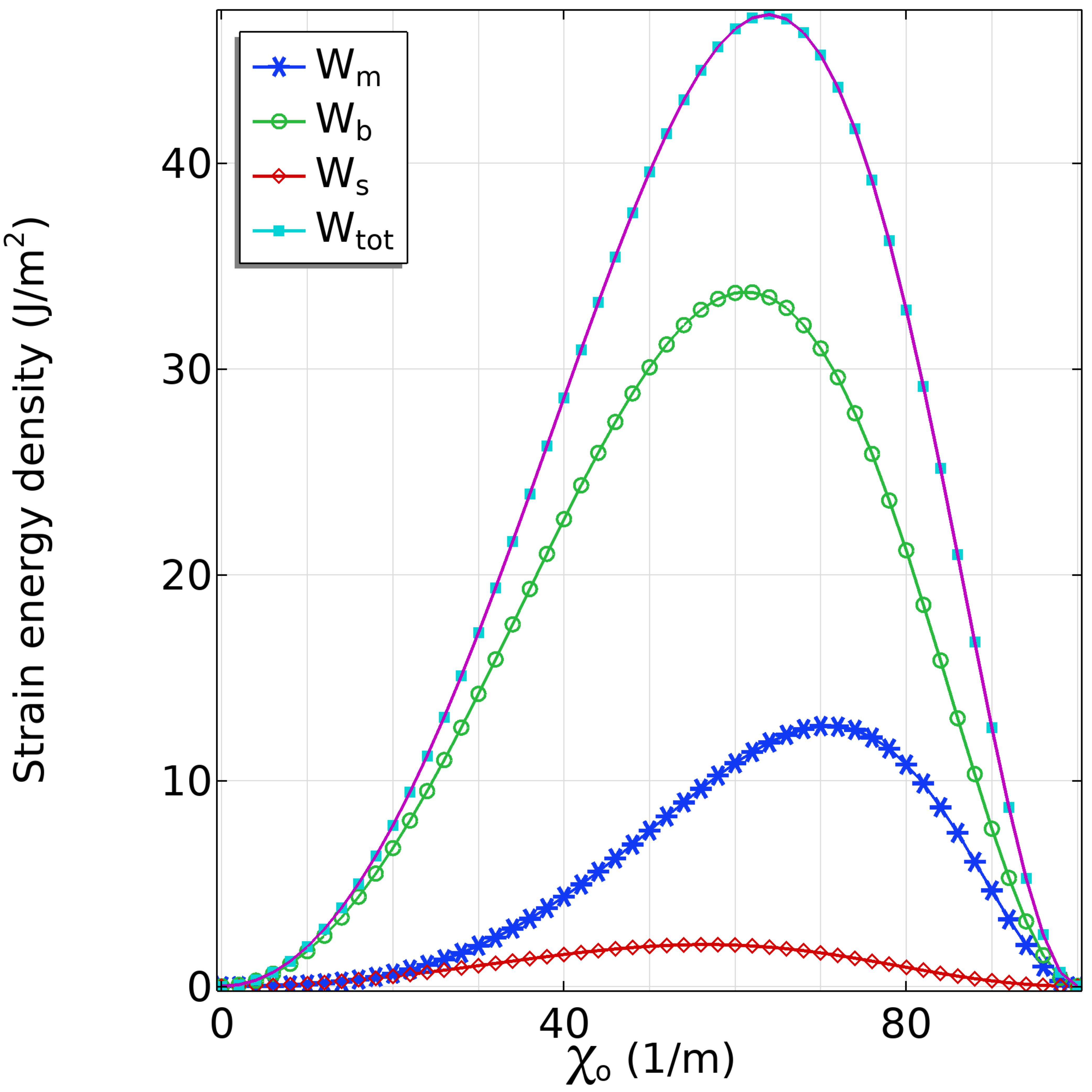}
\caption{Result for trigonometric fold-line. 
Left: Folding angle $\theta_{12}$ between the two folds; the inset shows the configuration at $\chi_o=100$ 1/m. Center: Current bending strain 
$\chi_{yy}$ versus $X/L$ along the center line of the shell for three different values of the target curvature $\chi_o$ evaluated at
the folding angles $\theta_{12}=30^o, 60^o, 90^o$. 
It might be noticed that current $\chi_{yy}$ differs from 
the target curvature $\chi_o$ as consequence of the incompatibility of the curvature assigned to the two folds, but contrary to the previous case with circular fold-line, at $\theta_{12}=90^o$ we have $\chi_{yy}=\chi_{o}$,
that is $\chi_{o}=100$ 1/m is a compatible target curvature.
Right: Elastic energies
versus target curvature $\chi_o$; it might be noticed that energies are not monotone with $\chi_o$; after increasing, they go back to zero
as consequence of compatibility a target curvature 
$\chi_{o}=100$ 1/m.}
  \label{fig:1F_SinN_theta_chi_ene}
\end{figure}

The main result is that for the circular fold line, any value of the
target curvature $\chi_o$ is incompatible and increases the energy of
the shell; see Fig. \ref{fig:1F_circ_theta_chi_ene}.
However, this energy is mostly due to bending ($W_b$), while the membrane contribution ($W_b$) is comparatively smaller. As shown in the inset, this is particularly true in the regime of moderate folding angles, the ones  smaller than $90^\circ$ where the geometric construction of Section~\ref{sec:geometric_approach} applies. For completeness, we also show the shear contribution $W_s$.
Conversely, for the trigonometric fold line there exists
a value of the target curvature, and hence of the folding angle, that yields a compatible folding, where also the bending energy vanishes.
For this geometry, this special curvature is $\chi_o=100$ 1/m and the corresponding folding angle is $110^\circ$, see Fig. \ref{fig:1F_SinN_theta_chi_ene}. To achieve this folding angle, an energy barrier with significant contributions from both membrane and bending energies needs to be overcome.

\begin{figure}
\center
\includegraphics[width=0.3\columnwidth]{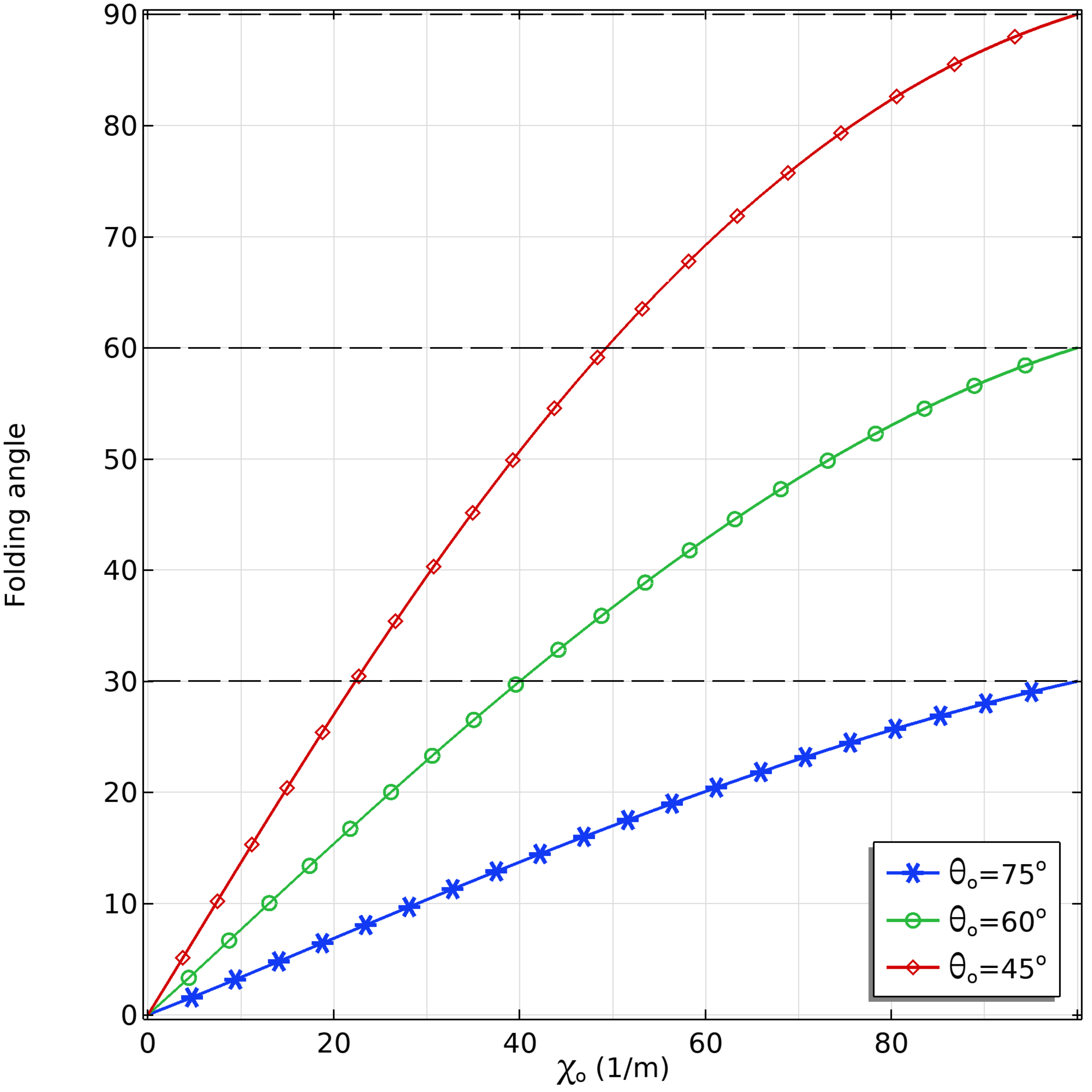}
\includegraphics[width=0.3\columnwidth]{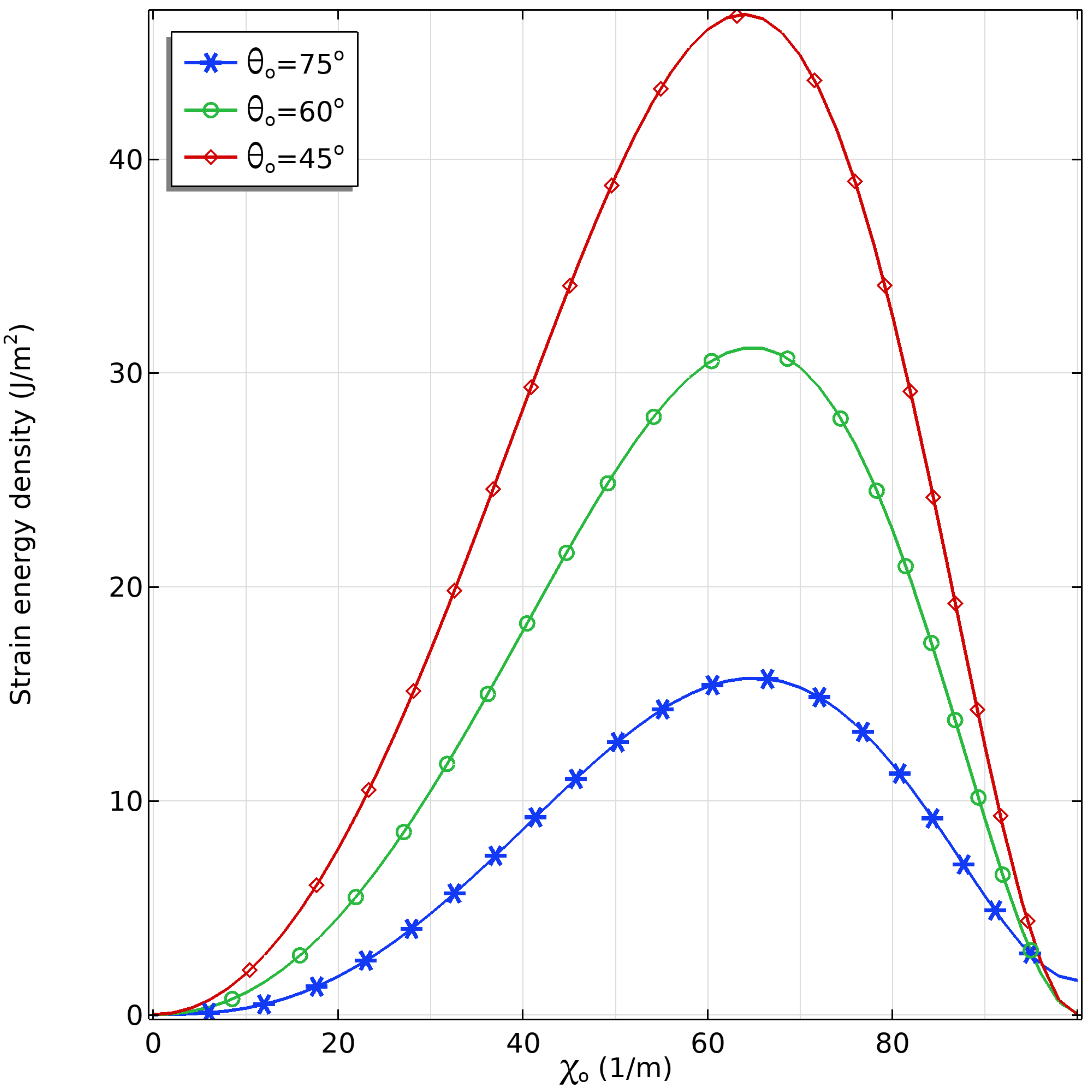}
\includegraphics[width=0.3\columnwidth]{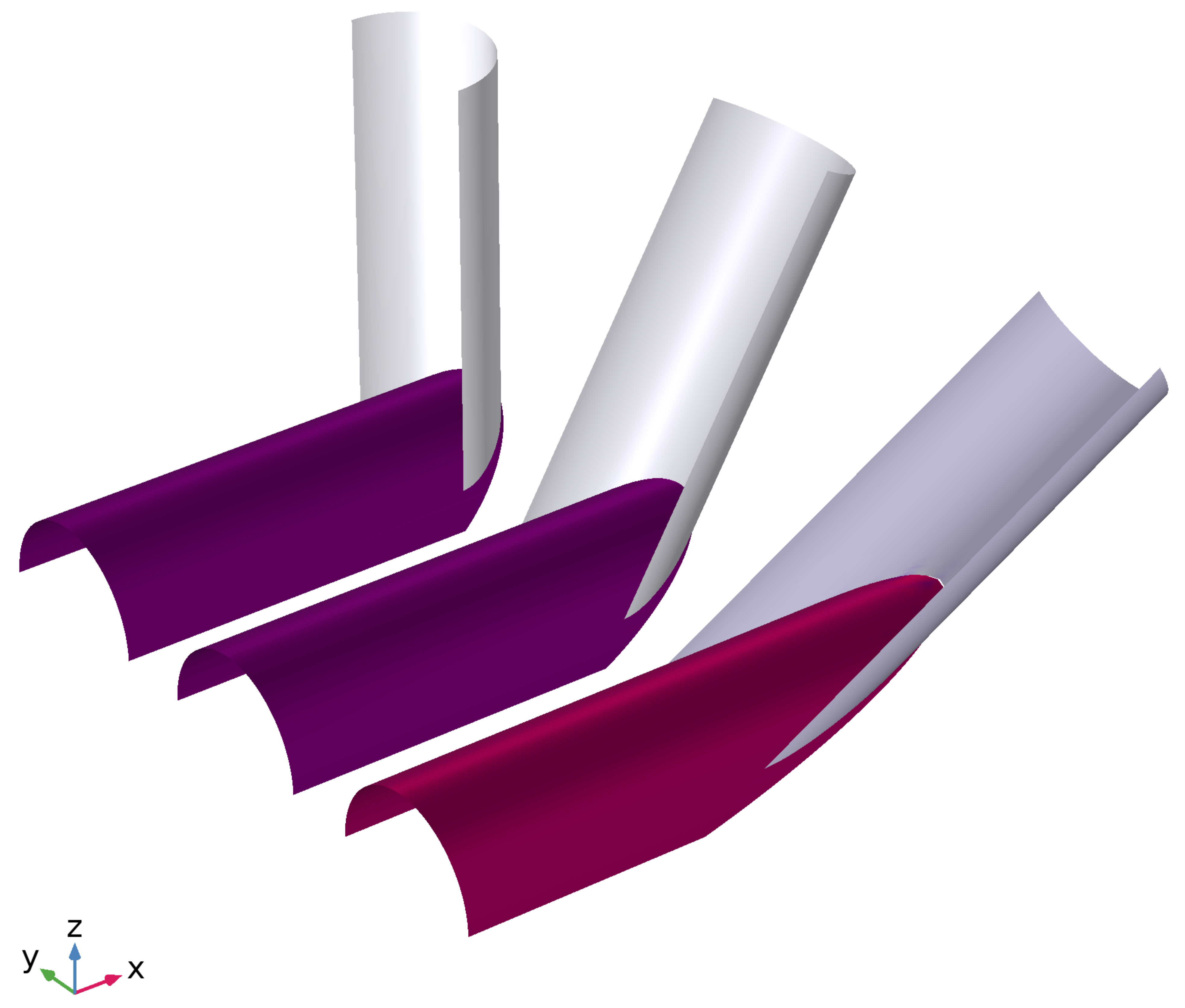}
\caption{Result for trigonometric fold-line with three different $\vartheta_o$. 
Left: Folding angle $\theta_{12}$ between the two folds; each fold-line has been designed
to have a compatible configuration at $\chi_o=100$ 1/m, and with 
folding angles $\theta_{12}=30^o, 60^o, 90^o$. 
Center: Elastic energies
versus target curvature $\chi_o$; it might be noticed that energies are not monotone with $\chi_o$; after increasing, they go back to zero as consequence of compatibility a target curvature 
$\chi_{o}=100$ 1/m.
Right: the three final configurations at $\chi_o=100$ 1/m.}
  \label{fig:1F_SinN_paraTheta_o}
\end{figure}

We note that using the trigonometric function, it is possible to design a fold line that
achieves a compatible configuration at a given
value of the target curvature $\chi_o$.
In Figure~\ref{fig:1F_SinN_paraTheta_o} we show
the results corresponding to three different fold line parametrized by $\vartheta_o=75^o, 60^o, 45^o,$ that
at $\chi_o=100$ 1/m, yields the folding angles $\theta_{12}=30^o, 60^o, 90^o$, respectively. 
For all three cases, the energy is not monotone with $\chi_o$; after increasing, it goes back to zero as consequence of the compatibility at 
$\chi_{o}=100$ 1/m.


\section{Interaction between folds: synchronous vs sequential folding in an array of 4 identical circular folds}
%
We consider a flat rectangular shell $\mathcal{S}=L\times W$ 
split in four folds by three circular fold-lines
We look for the configuration
associated to given values of the target curvatures 
$\chi_{oj}$ in each fold $j=1,\ldots,4$. 
In particular, we assign a sequence of $n$
quadruplets $X_{oi}=(\chi_{01},\ldots,\chi_{o4})_i$,
with $i=1,\ldots,n$ and $\chi_{oi}$ ranging 
in $(0, \pm\chi_{m})$ (1/m),
and we solve for the corresponding configuration $\mathcal{C}_i$;
we denote with $\theta_{hk}=2\,\alpha_{hk}$ the 
folding angle between fold $h$ and $k$.
Two different sequences are used:
\begin{itemize}
\item Sync (Synchronized): 
$X_{oi}=(-\chi_{01},\chi_{01},-\chi_{01},\chi_{01})_i$.
The alternating sign $\pm$ of the target curvatures in 
the adjacent folds yields an almost compatible
deformation; the change of configuration is smooth, and 
associated to a low increase of elastic energy.
\item Seq (Sequential): the whole sequence $X_{oi}$ is split
into three sub sequences.
\be
\begin{array}{lll}
X_{oi}=(-\chi_{01},\chi_{01},\chi_{01},\chi_{01})_i,\, 
& \textrm{with }\chi_{01}=0,\ldots,\chi_{m} ,\,\,
& i=1,\ldots,n_1, \\[2mm]
X_{oi}=(-\chi_{m},\chi_{m},\chi_{03},-\chi_{03})_i,\,
& \textrm{with } \chi_{03}=\chi_{m},\ldots,-\chi_{m},\,\,
&  i=n_1+1,\ldots,n_2,\\[2mm]
X_{oi}=(-\chi_{m},\chi_{m},-\chi_{m},\chi_{04})_i,\,
& \textrm{with } \chi_{04}=-\chi_{m},\ldots,\chi_{m},\,\,
& i=n_2+1,\ldots,n_3, \\[2mm]
\end{array}
\ee
The first sequence changes only the angle $\theta_{12}$
between the first fold and the next three. The second one maintains $\theta_{12}$ constant, and only changes $\theta_{23}$; the final sequence maintains both 
$\theta_{12}$ and $\theta_{23}$ constant, and changes
$\theta_{34}$. The change of configuration might exhibt
jumps, and elastic energy is not monotone with the solution
number $i$.
\end{itemize}

\begin{figure}[htbp]
\centering
\includegraphics[width=0.32\columnwidth]{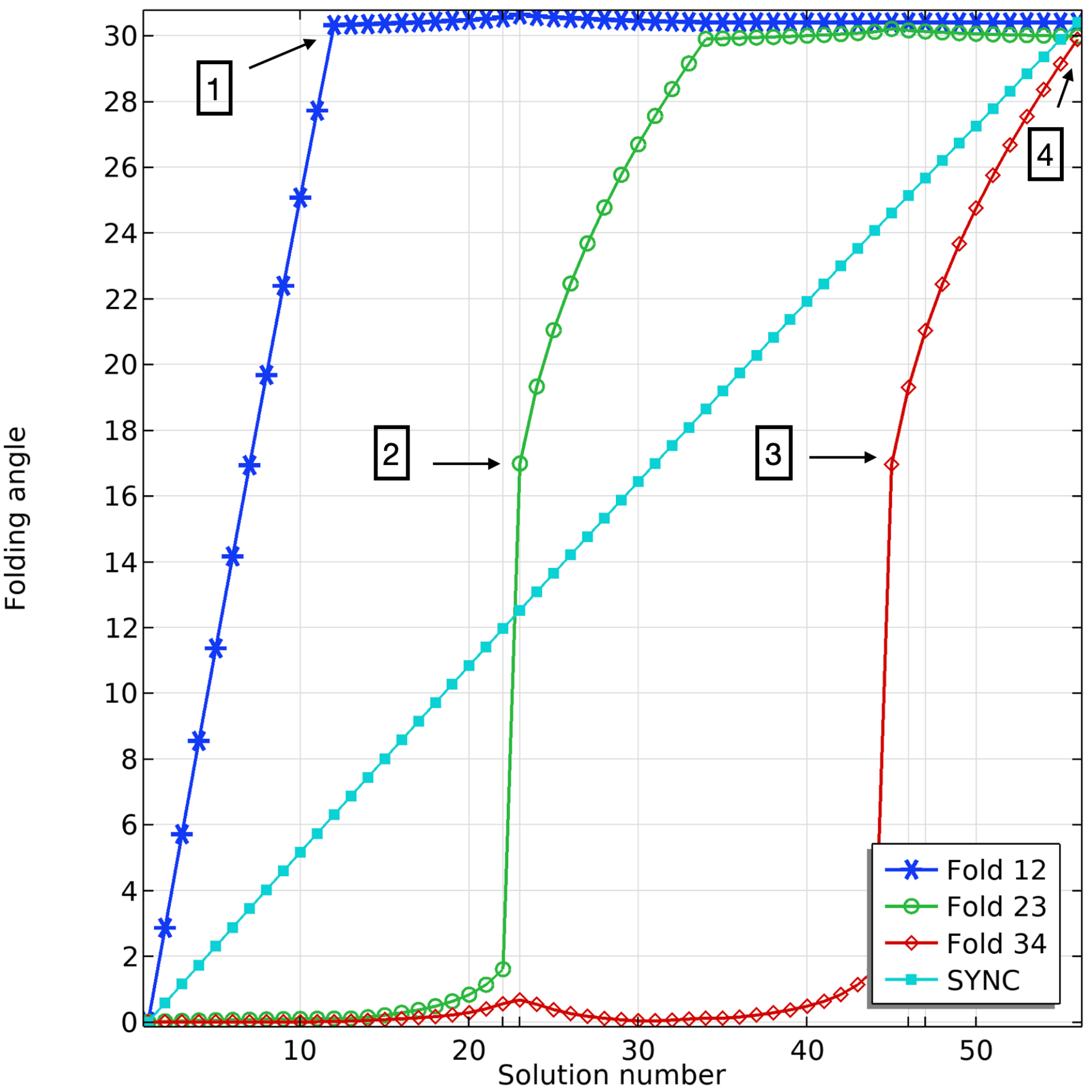}
\includegraphics[width=0.32\columnwidth]{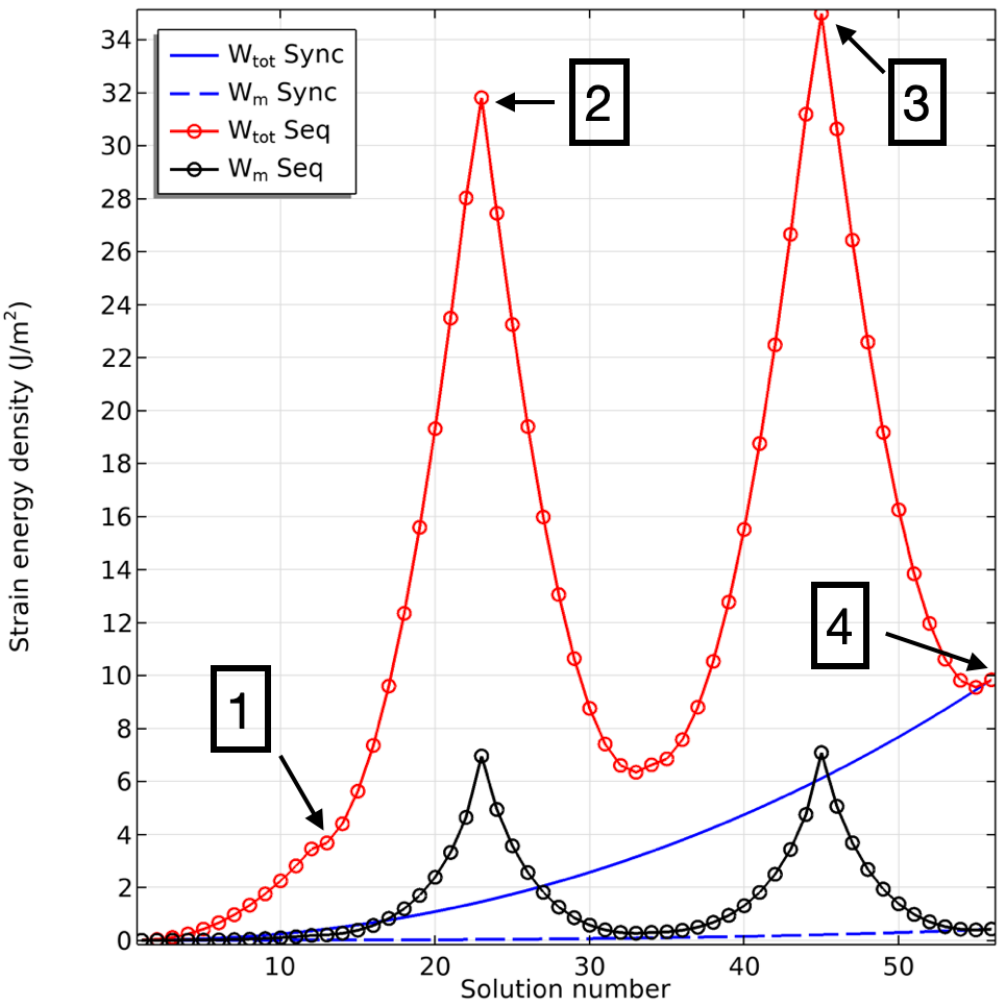}
\includegraphics[width=0.32\columnwidth]{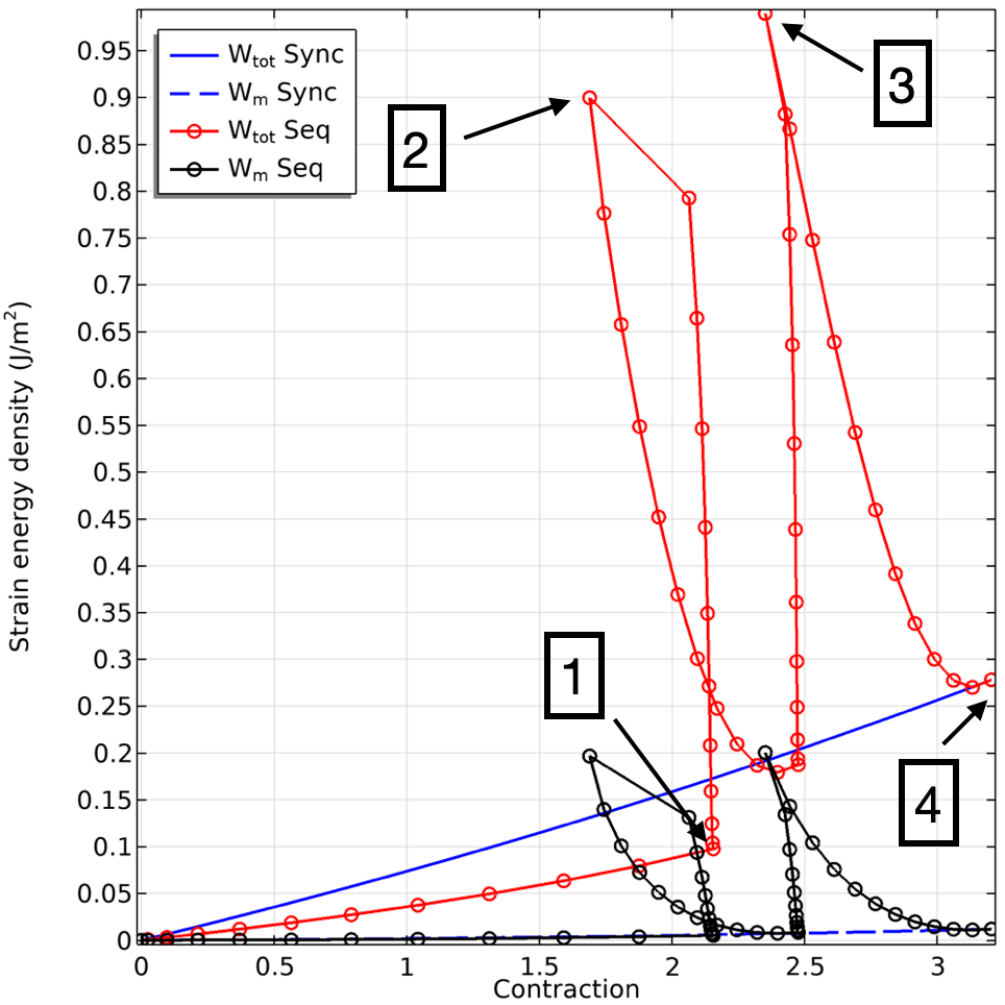}
\caption{Elastic energy versus solution number for the synchronized and the sequential folding. The four numbered points correspond to the configurations in Fig.~\ref{fig:circ_3F_30deg}.} 
\end{figure}

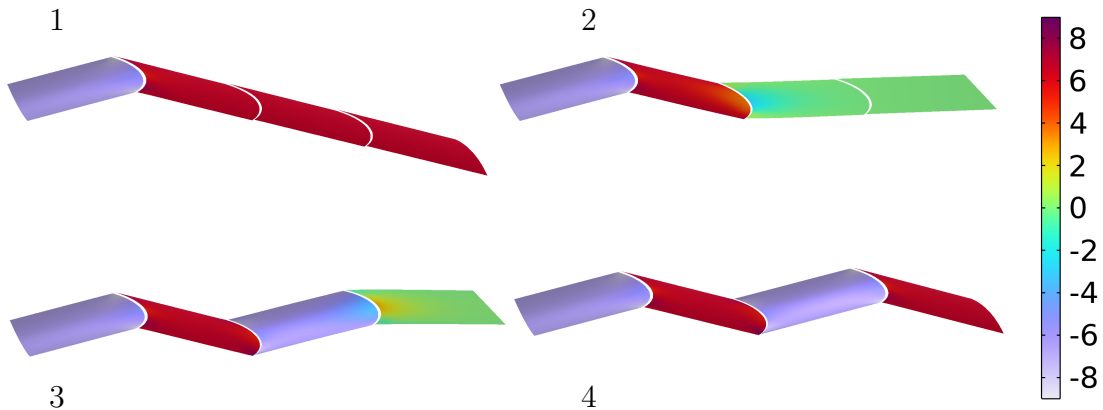
\begin{figure}[htbp]
\centering
\begin{tikzpicture}[scale=1]
\node[inner sep=0pt] at (7.5,3)
{\includegraphics[width=1\columnwidth]{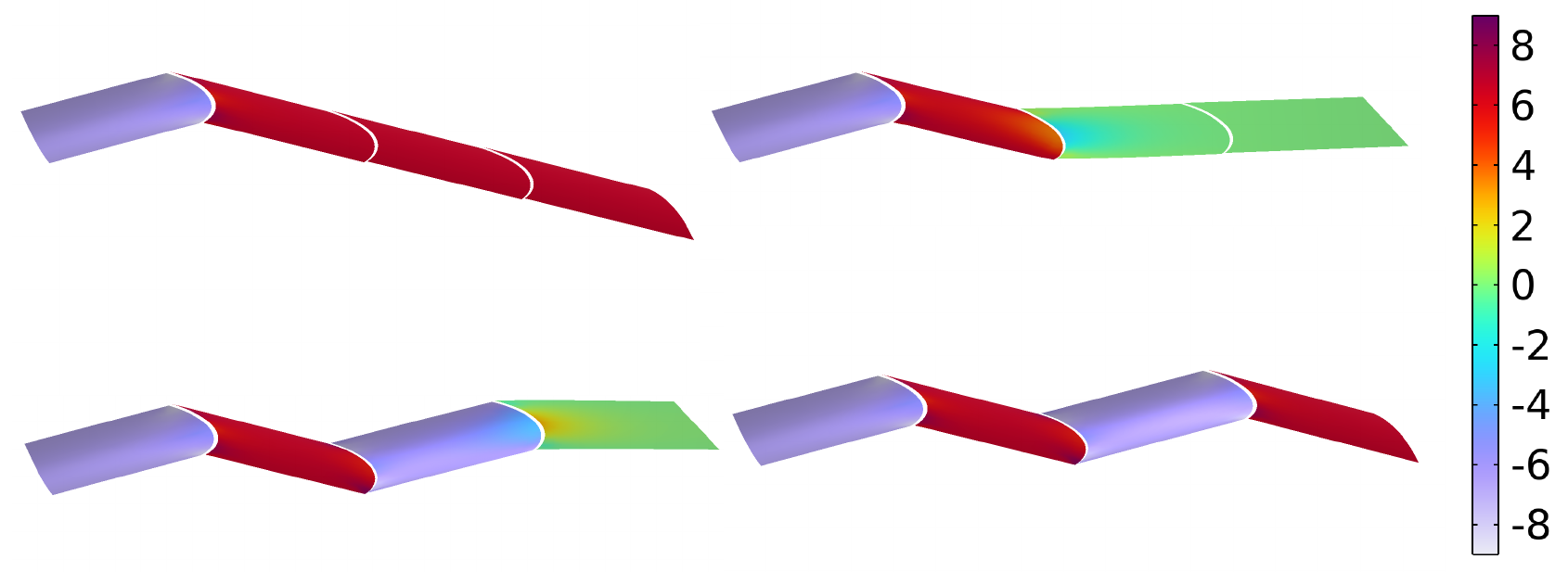}};
%
\draw (1,5.5) node {\large $1$};
\draw (8,5.5) node {\large $2$};
\draw (1,.5) node {\large $3$};
\draw (8,.5) node {\large $4$};
\end{tikzpicture}
\caption{Four different configurations of the shell made of four folds, undergoing a sequential folding; circular fold-lines. Color legend shows the bending strain $\chi_{yy}$ in the transversal direction (1/m).} 
\label{fig:circ_3F_30deg}
\end{figure}


\section{Interaction between folds: fold pairs of circular and trigonometric arcs with different radii or heights}

Following the modeling approach described
in Sec.~\ref{sec:shell}, 
we compare the behavior of flat rectangular shells $\mathcal{S}=L\times W$ 
split in three folds by different pair of fold-lines:
1) two circular arcs, with different curvature radii $R_o$; 2) two trigonometric function, with different amplitude.
The geometries are described in Sec. \ref{sec:geometries}.
The shells are not loaded, and have compatible constraints, that is
the deformations they exhibit do not generate reaction forces; the only input is the 
target curvature $\chi_o$, see
(\ref{kappa_o}).

For both shells, we solve for the configurations $\mathcal{C}_i$
corresponding to a sequence of $n$ triplets $X_{oi}=(\chi_{oi},-\chi_{oi},\chi_{oi})$,
where $\chi_{oi}$ are the target curvatures assigned to the first and third fold
and  $-\chi_{oi}$ the ones of the middle fold;
$\chi_{oi}$ ranges in $(0, \chi_{m})$ (1/m).

We discuss the results showing: 1) the folding angles 
$\theta_{12}=2\,\alpha_{12}$ between the first and the second fold, and $\theta_{23}=2\,\alpha_{23}$
between the second and the third fold,

2) the actual curvature $\chi_{yy}$ of the deformed configuration $\mathcal{C}$
evaluated along the center line;
3) the elastic energy density (energy per unit area).

\begin{figure}
\center
\includegraphics[width=0.32\columnwidth]{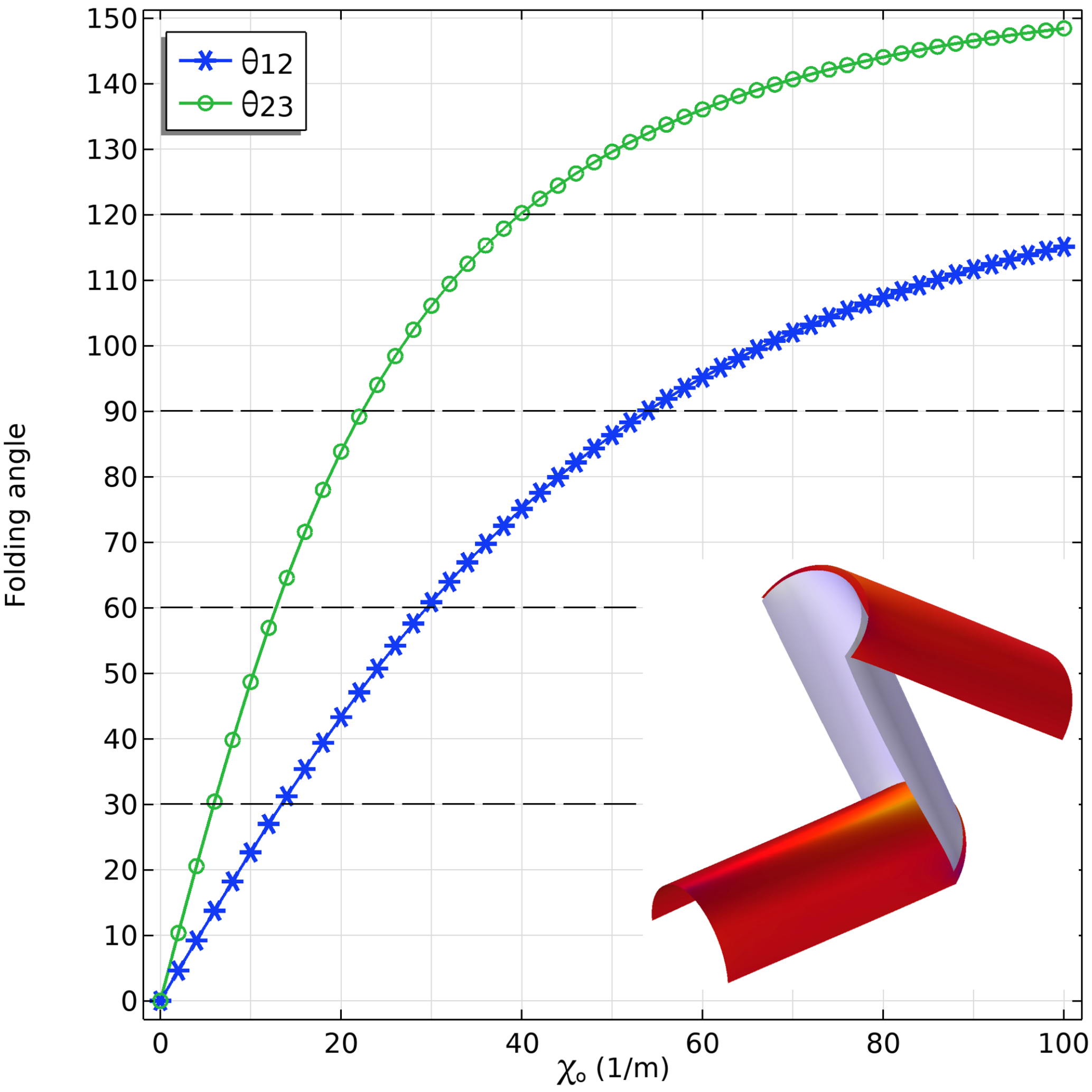}
\includegraphics[width=0.32\columnwidth]{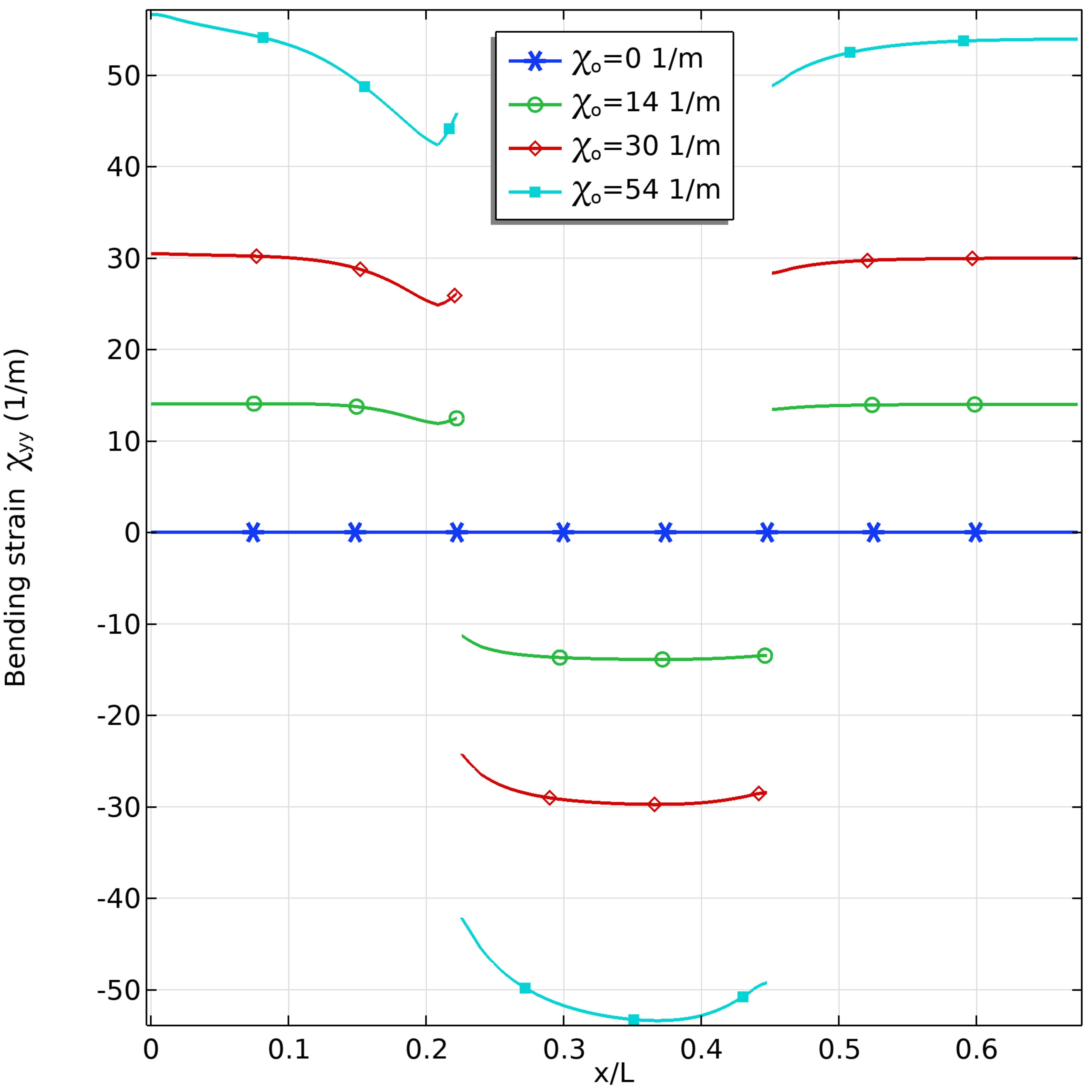}
\includegraphics[width=0.32\columnwidth]{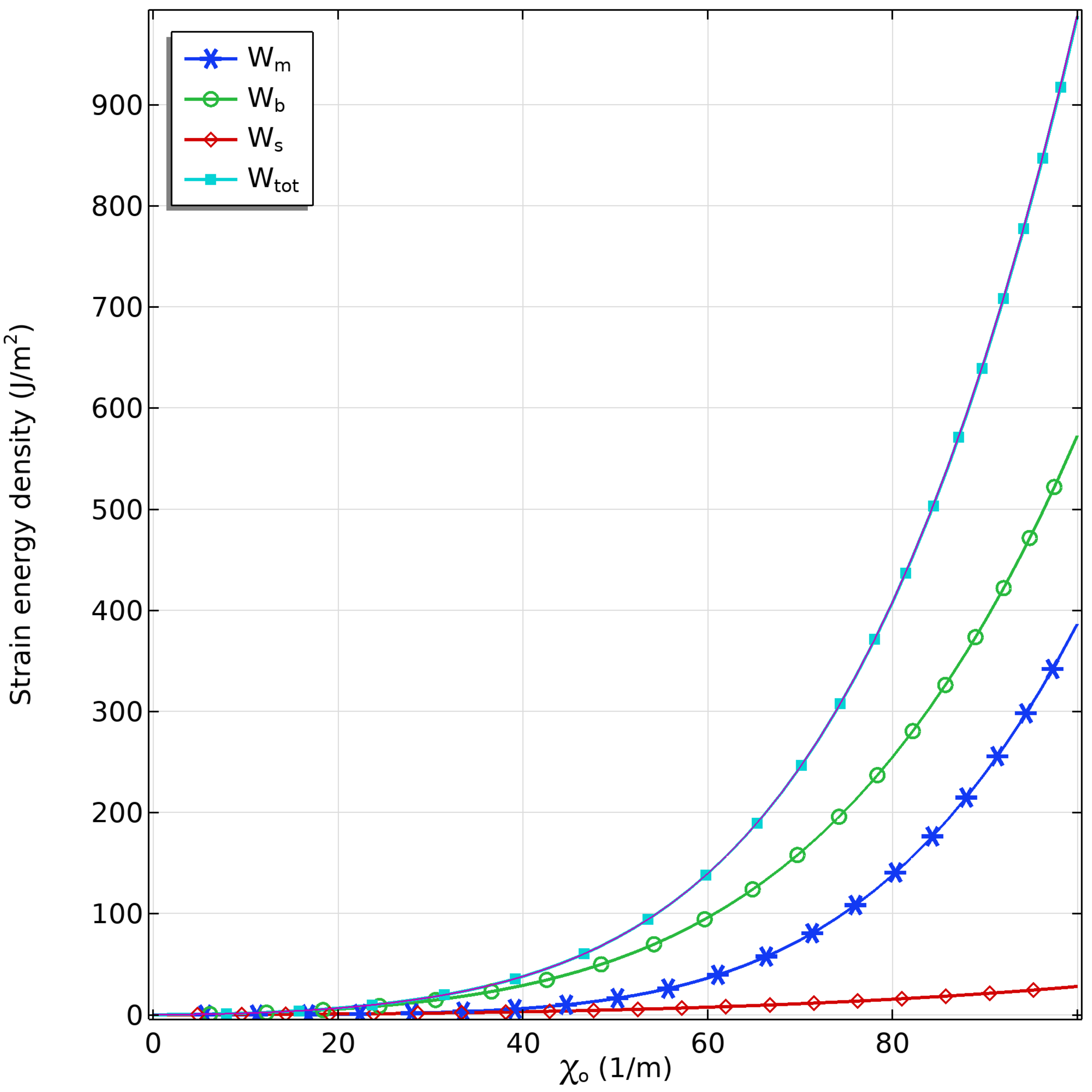}
\caption{Result for two different circular fold-lines. 
Left: Folding angle $\theta_{12}$ and
$\theta_{23}$ between the two pairs of adjacent folds; we note that for any $\chi_o$ there correspond different folding angles.
The inset shows the configuration at $\chi_o=100$ 1/m. Center: Current bending strain 
$\chi_{yy}$ versus $X/L$ along the center line of the shell for three different values of the target curvature $\chi_o$ evaluated at
the folding angles $\theta_{12}=30^o, 60^o, 90^o$. 
It might be noticed that current $\chi_{yy}$ differs from 
the target curvature $\chi_o$ as consequence of the incompatibility of the curvature assigned to the three folds. The difference between $\chi_{yy}$ and $\chi_{o}$ increases with the folding angle.
Right: Elastic energies
versus target curvature $\chi_o$; it might be noticed that energies increases with $\chi_o$, a consequence of the aforementioned
incompatibility of $\chi_o$.}
  \label{fig:2F_circ_theta_chi_ene}
\end{figure}

\begin{figure}
\center
\includegraphics[width=0.3\columnwidth]{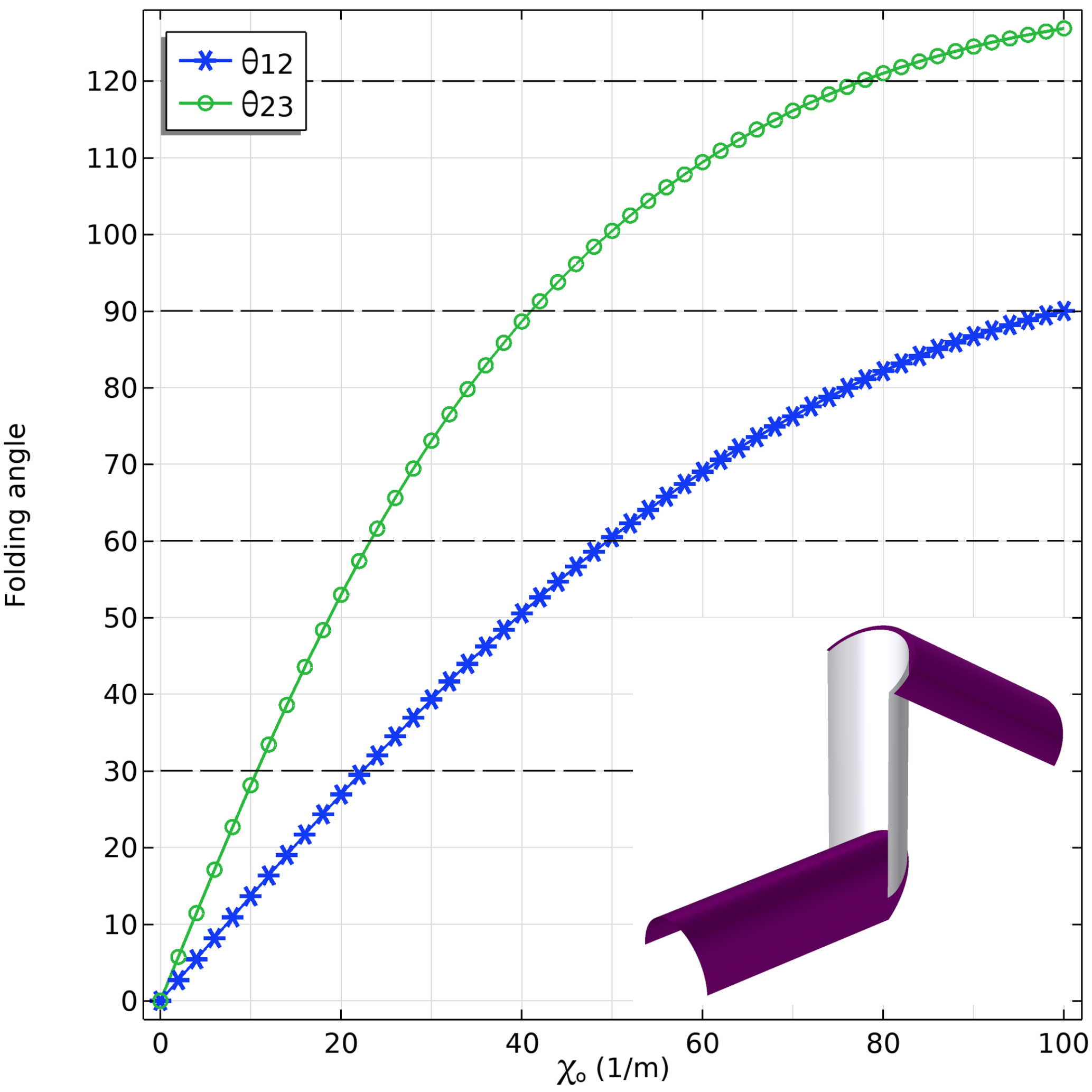}
\includegraphics[width=0.3\columnwidth]{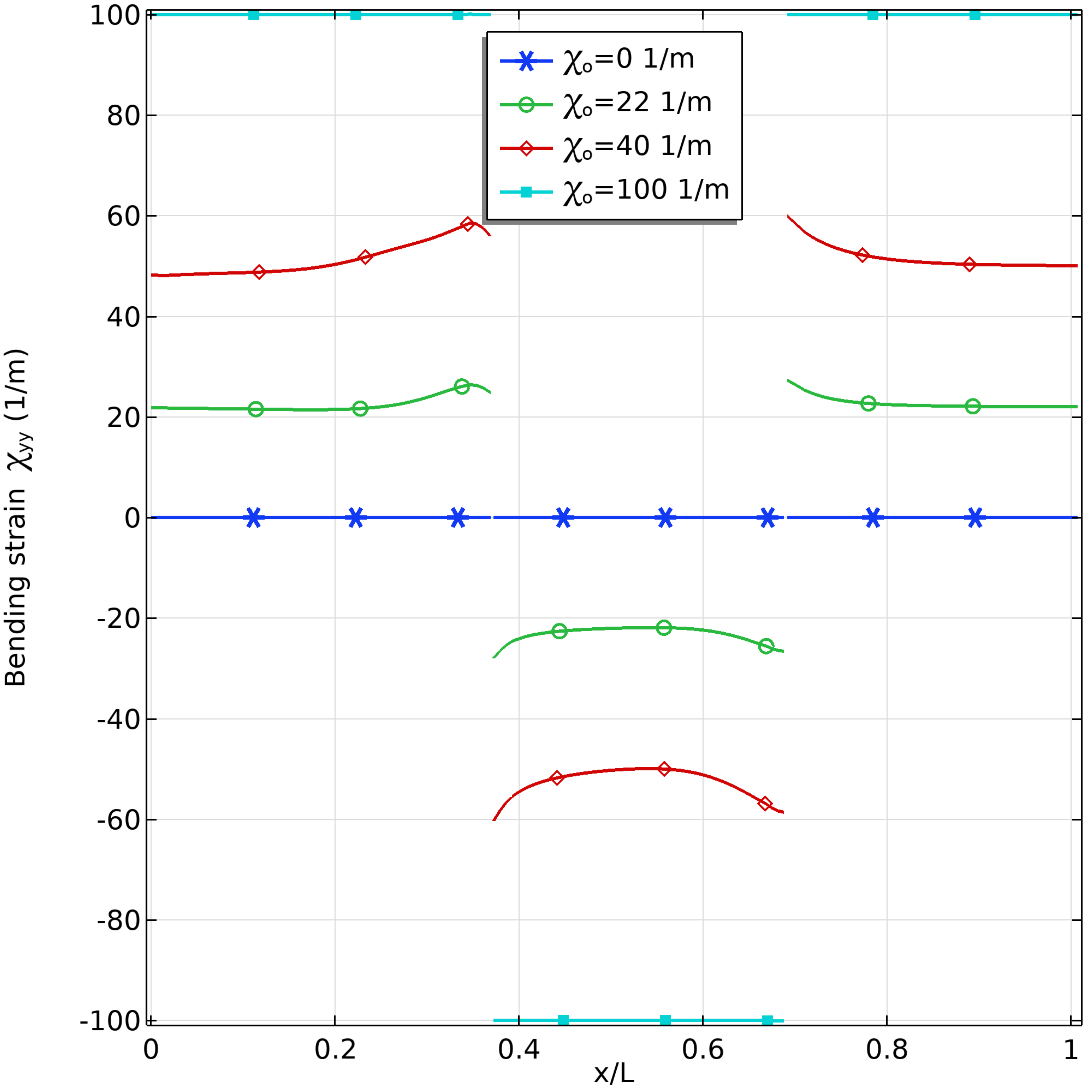}
\includegraphics[width=0.3\columnwidth]{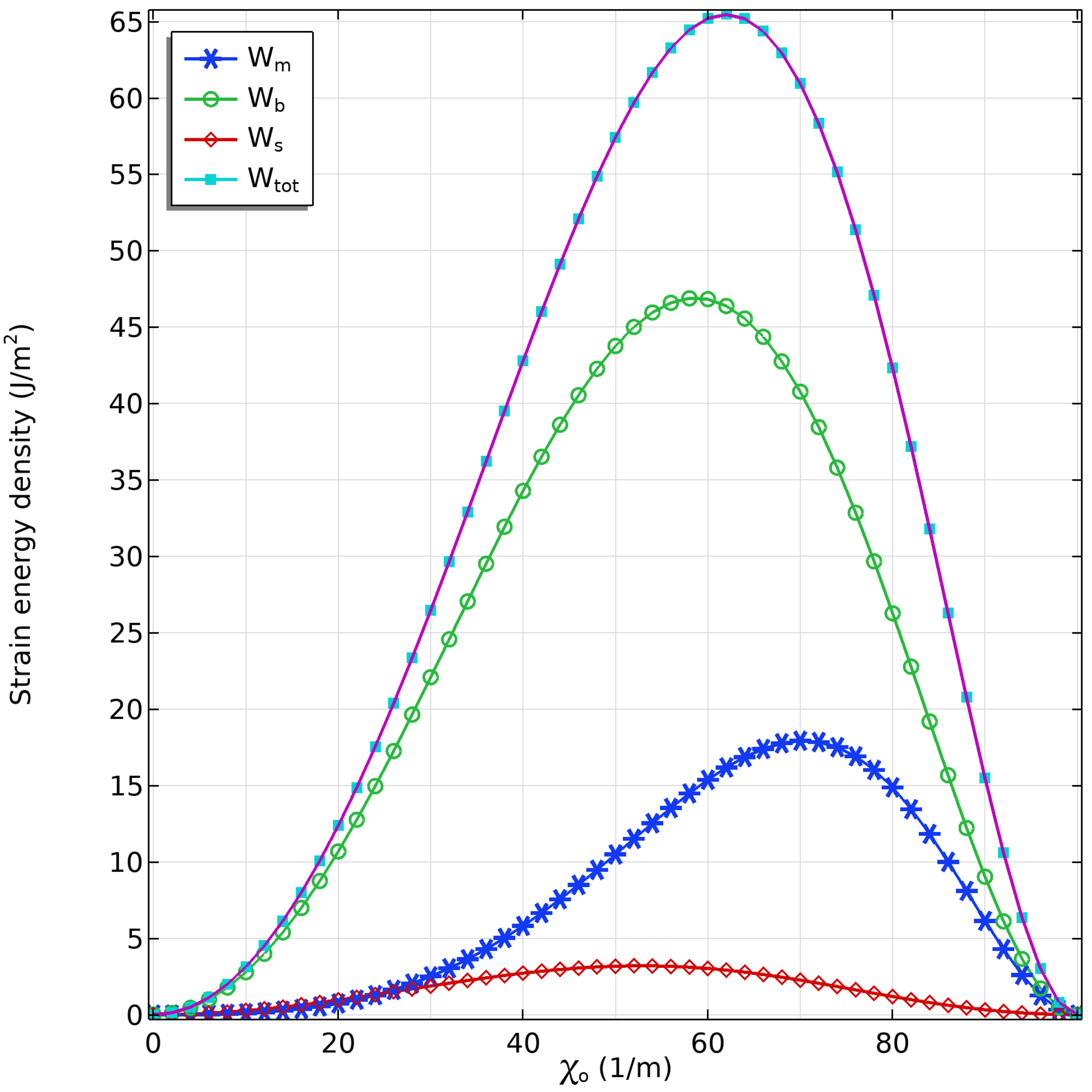}
\caption{Result for two different 
trigonometric fold-lines. 
Left: Folding angle $\theta_{12}$ and
$\theta_{23}$ between the two pair of adjacent folds; we note that for any $\chi_o$ there correspond different folding angles.
The inset shows the configuration at $\chi_o=100$ 1/m.  Center: Current bending strain 
$\chi_{yy}$ versus $X/L$ along the center line of the shell for three different values of the target curvature $\chi_o$ evaluated at
the folding angles $\theta_{12}=30^o, 60^o, 90^o$. 
The current $\chi_{yy}$ differs from 
the target curvature $\chi_o$ as consequence of the incompatibility of the curvature assigned to the three folds, but contrary to the previous case with circular fold-lines, at $\theta_{12}=90^o$ we have $\chi_{yy}=\chi_{o}$,
that is $\chi_{o}=100$ 1/m is a compatible target curvature for both folds.
Right: Elastic energies
versus target curvature $\chi_o$; the energies are not monotone with $\chi_o$, and after increasing, they go back to zero
as consequence of compatibility at 
$\chi_{o}=100$ 1/m.}
  \label{fig:2F_SinN_theta_chi_ene}
\end{figure}

The main result is that both geometries exhibit
two different folding angles, that is
$\theta_{12}\ne\theta_{23}$.
Moreover, as already observed, for the circular fold line, any value of the
target curvature $\chi_o$ is incompatible and increases the energy of the shell; see Fig. \ref{fig:2F_circ_theta_chi_ene}.
Conversely, for the trigonometric fold-line there exists
a value of the target curvature that yields a compatible folding:
for this geometry it is $\chi_o=100$ 1/m, see Fig. \ref{fig:2F_SinN_theta_chi_ene}.

\subsection{Mechanics modeling}
%
\subsubsection{The 3D model}\label{sec:3D_model}
%
The reference configuration of the 3D solid is described by
$\mb{X}=(X,Y,Z)$, where $(X,Y,Z)$ are the material coordinates of $\mathcal{V}$.
The current positions $x=(x,y,z)$ of material points $X$ are given by 
$x=X+\mb{u}(X)$, with $\mb{u}$ the displacement. The strain $\mb{E}$ and stress $\mb{S}$
are defined by
\be
\mb{E}=\frac12\,(\nabla\mb{u}+\nabla\mb{u}^T +\nabla\mb{u}^T\,\nabla\mb{u}),
\quad
\mb{S}=\mathbb{C}\,\mb{E}_e,
\quad\textrm{and } \mb{E}_e=\mb{E}-\mb{E}_o, 
\ee
where $\mathbb{C}$ is the elastic tensor corresponding to an isotropic
material, and $\mb{E}_e$ is the elastic strain.
The target strain $\mb{E}_o$ is defined by
\be\label{lambda_o}
\mb{E}_o=\lambda_\| \,\mb{e}_2\otimes\mb{e}_2 
        + \lambda_\bot\,(\mb{e}_1\otimes\mb{e}_1+\mb{e}_3\otimes\mb{e}_3),
\quad\textrm{with }
\lambda_\|=1+\varepsilon,\,\lambda_\bot=1.
\ee
The target strain $\mb{E}_o$ yields a distortion along the $Y$ direction
which induces a bending in the plane $(Y,Z)$.
The elastic energy $\Psi$ is given by
\be
\Psi=\frac12\,\mathbb{C}\,\mb{E}_e\cdot\mb{E}_e
\ee
Finally, assuming no loads, 
global energy minimization yields to the following weak equation:
\be\label{3D_weak}
\int_{\mathcal{V}} \, \mb{S}  \cdot \tilde{\mb{E}}_e \,dV = 0\,,
                  \quad
\forall\, \tilde{\mb{u}}\,.
\ee
We solve (\ref{3D_weak}) using as input a sequence of target stretches
$\lambda_\|=1,\ldots,\lambda_{max}$, with a minimal set of kinematical constraints that rule out rigid motions without inducing reactive forces.
%
\subsubsection{The Shell model}\label{sec:shell_model}
%
The reference configuration of the shell-like region is given by
\be
\mathbb{R}^3\ni\mb{X}=\mb{X}(\xi_1,\xi_2,\xi_3)
      =\mb{s}(\xi_1,\xi_2) +\xi_3\,\mb{N}(\xi_1,\xi_2)\,.
\ee
where $\mb{s}:\mathbb{R}^2\to\mathbb{R}^3$  is the reference shell surface,
$\mb{N}$ the shell normal, and $\xi_i$ are 
the local coordinates on $\mathcal{S}$. 
The coordinate $\xi_3$ in the normal direction ranges in
$(-H/2,+H/2)$.
The state variables of the shell model
are the displacement $\mb{u}$,
yielding the current position of the reference shell surface
$\mathcal{S}$, and the vector $\mb{a}$ that gives the current
normal $\mb{n}=\mb{N}+\mb{a}$.
The current configuration of the 3D shell-like
region is given by
\be
\mb{x}=\mb{x}(\xi_1,\xi_2,\xi_3)
      =\mb{s}(\xi_1,\xi_2)+\mb{u}(\xi_1,\xi_2)
      +\xi_3\,\mb{n}(\xi_1,\xi_2)\,.
\ee
Let the indices $\alpha$ and $\beta$ range in $1,2$. 
The in plane Green-Lagrange strains are measured by
\be
\epsilon_{\alpha\,\beta}=\frac12\,\left(
 \frac{\partial \mb{x}}{\partial \xi_\alpha}
\cdot \frac{\partial \mb{x}}{\partial \xi_\beta}
-\frac{\partial \mb{X}}{\partial \xi_\alpha}
\cdot \frac{\partial \mb{X}}{\partial \xi_\beta}\right)
=\gamma_{\alpha\,\beta}+\xi_3\,\chi_{\alpha\,\beta}
  +O(\xi_3^2)\,\dot{=}\,\gamma_{\alpha\,\beta}+\xi_3\,\chi_{\alpha\,\beta}\,.
\ee
The transverse shear is given by 
\be
\epsilon_{3 \alpha}=\epsilon_{\alpha 3}
 =\frac12\,\left(
 \frac{\partial \mb{x}}{\partial \xi_\alpha}\cdot \mb{n}
-\frac{\partial \mb{X}}{\partial \xi_\alpha}\cdot \mb{N}\right)
=\zeta_\alpha+O(\xi_3)\,\dot{=}\,\zeta_\alpha\,.
\ee
Note that the contributions of order $O(\xi_3^2)$ from 
$\epsilon_{\alpha\,\beta}$, and order $O(\xi_3)$ from
$\epsilon_{3 \alpha}$ have been considered negligible, as well as
$\epsilon_{33}$. 
The remaining contributions are 
called \emph{membrane strain} $\gamma_{\alpha\,\beta}$,
\emph{bending} $\chi_{\alpha\,\beta}$,
and \emph{shear} $\zeta_{\alpha}$.
The membrane stress $\sigma_m$, the bending stress $\sigma_b$ 
and the shear stress $\sigma_s$ are given by
\be
\sigma_m=\mb{D}\,(\gamma-\gamma_o)\,,
\quad
\sigma_b=\frac{H}{2}\,\mb{D}\,(\chi-\chi_o)\,,
\quad
\sigma_s=\frac{5}{6}\,2\,G\,(\zeta-\zeta_o)\,,
\ee
where $\mb{D}$ is the 2D plane stress constitutive matrix corresponding
to the 3D elastic tensor $\mathbb{C}$, 
$G$ the transverse shear moduli,
and $\gamma_o$, $\chi_o$, $\zeta_o$ are the target or target strains.
The current stress $\sigma$ in the 3D shell-like region is recovered by 
the formula $\sigma=\sigma_m+z\,\sigma_b$, with $z$ is a 
non dimensional coordinate ranging from $-1$ (bottom surface) to $1$ 
(top surface). 
The shell membrane-force $\mb{N}_m$, bending moment $\mb{M}_b$, 
and shear force $\mb{Q}_s$ are computed from the stresses as follows
\be
\mb{N}_m = H\,\sigma_m, \quad
\mb{M}_b = \frac{H^2}{6}\,\sigma_b, \quad
\mb{Q}_s = H\,\sigma_s\,.
\ee
The previous constitutive relations might be rewritten in matrix form
as:
\be\label{costi}
\left[\begin{array}{c} \mb{N}_m \\[2mm] 
                       \mb{M}_b \\[2mm]
                       \mb{Q}_s  
     \end{array}\right] =
\left[\begin{array}{ccc} \mb{D}_A & \mb{D}_B & 0 \\[2mm]
                         \mb{D}_B & \mb{D}_D & 0 \\[2mm]
                            0     &     0    & \mb{D}_{AS}
    \end{array}\right]
\,
\left[\begin{array}{c} \gamma_e \\[2mm]
                       \chi_e     \\[2mm]
                       \zeta_e   
    \end{array}\right],
\textrm{ with }
    \gamma_e=\gamma-\gamma_o,\,
    \chi_e=\chi-\chi_o,\,
    \zeta_e =\zeta-\zeta_o\,.
\ee
The stiffness matrices in \eqref{costi} have following IS units:
\be
[\mb{D}_A]=\rm{N/m}, \quad
[\mb{D}_B]=\rm{N}, \quad
[\mb{D}_D]=\rm{N m}, \quad
[\mb{D}_{AS}]=\rm{N/m}\,.
\ee
The elastic energy $\psi_{ela}$ 
can be represented as the sum of the membrane
$\psi_m$, the bending $\psi_b$, and the shear energy $\psi_s$ as follows:
\be\label{shell_energy}
\begin{array}{lcl}
\psi_{ela}&=&\psi_m+\psi_b+\psi_s \\[3mm]
          &=&\dfrac12\,(\mb{D}_A\,\gamma_e+\mb{D}_B\,\chi_e)\cdot\gamma_e
           + \dfrac12\,(\mb{D}_B\,\gamma_e+\mb{D}_D\,\chi_e)\cdot\chi_e
          + \dfrac12\, \mb{D}_{AS}\,\zeta_e\cdot\zeta_e \\[3mm]
    &=& \dfrac12\,\mb{N}_m\cdot\gamma_e 
     +  \dfrac12\,\mb{M}_b\cdot\chi_e 
     +  \dfrac12\,\mb{Q}_s\cdot\zeta_e.       
\end{array}
\ee
Finally, assuming no loads, 
global energy minimization yields the following weak equations:
\be\label{shell_weak}
\int_{\mathcal{S}} \left(\, \mb{N}_m  \cdot \tilde{\gamma} 
                  +\mb{M}_b \cdot \tilde{\chi}
                  + \mb{Q}_s  \cdot \tilde{\zeta}\,\right)\,dA = 0\,,
                  \quad
\forall\, \tilde{\gamma}, \, \tilde{\chi},\, \tilde{\zeta}\,.
\ee
%
%
%
\subsection{Geometries}\label{sec:geometries}
\begin{figure}
\center
\includegraphics[width=0.3\columnwidth]{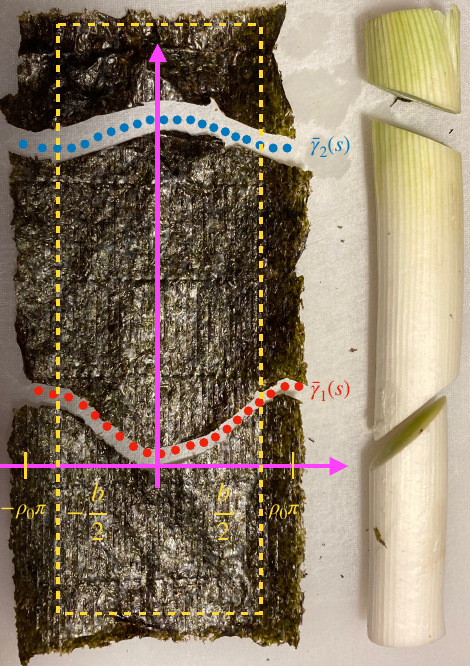}
\caption{Illustration of the unfolding of an oblique section of a right cylinder onto a plane, leading to trigonometric fold lines.}
  \label{fig:AlgaNori_curves}
\end{figure}

%
We list the shell geometries used to implement and solve our problems.
The footprint of each geometry is defined by a rectangle 
$L\times W$; the shell thickness is $H=L_o/100$.
The width $W$ is defined by: $W=\pi\,\rho_o$ for the circular fold-line;
$W=2\,\varphi\,\rho_o$ for the trigonometric fold line.
Finally, $\rho_o=1$, and $h_o=2\,\rho_o\,\tan(\vartheta_o)$,  $\vartheta_o=40^\circ$. The trigonometric fold line is obtained from the oblique section of a right cylinder with a plane (radius $\rho_0$, angle $\vartheta_o$), by unfolding the cylinder onto a plane, see Fig.~\ref{fig:AlgaNori_curves}.
For each fold line $i$, we have: $c_i$ = base point; $X_i$ and $Y_i$ parametric curve; $s\in(-\varphi,\varphi)$ abscissa of the curve.

\par\noindent
\emph{1 Fold:} $L=L_o=2\,W$ or $3\,W$; $R_o=L_o/4$.
\par\noindent
\begin{itemize}
\item Circular arc; $\varphi=\arcsin(2/3)$.
\be
c_1=(L/4,0), \quad 
\left\{\begin{array}{l}
X_1=R_o\,\sin(s+\pi/2) \\[2mm]
Y_1=R_o\,\cos(s+\pi/2) 
\end{array}\right.
\ee
\item Trigonometric function; $\varphi=\pi/2$.
\be
c_1=(L/2-h_o/4\,\sin(\varphi),0), \quad  \quad
\left\{\begin{array}{l}
X_1=h_o/2\,\sin(s+\pi/2) \\[2mm]
Y_1= \rho_o\,s 
\end{array}\right.
\ee
\par
\end{itemize}
\par\noindent
\emph{2 Folds:} $L=3\,L_o/2$, and $L_o=3\,W$, $R_o=L_o/4$.
\par\noindent
\begin{itemize}
\item Circular arc, different folds; $\varphi_1=\arcsin(2/3)$,
$\varphi_2=\arcsin(1/3)$,
\be\label{2F_arc_different}
\begin{array}{l}
c_1=(L_o/2,0), \quad 
\left\{\begin{array}{l}
X_1=R_o\,\sin(s+\pi/2) \\[2mm]
Y_1=R_o\,\cos(s+\pi/2) 
\end{array}\right., 
\quad
c_2=(L_o,0),\, 
\left\{\begin{array}{l}
X_2=2\,R_o\,\sin(s+\pi/2) \\[2mm]
Y_2=2\,R_o\,\cos(s+\pi/2) 
\end{array}\right., \,
\end{array}
\ee
\item Trigonometric function, different folds; $\varphi=\arcsin(2/3)$.

\be
\begin{array}{l}
c_1=(L_o/2-h_o/4\,\sin(\phi),0),\, 
\left\{\begin{array}{l}
X_1=h_o/2\,\sin(s+\pi/2) \\[2mm]
Y_1= \rho_o\,s 
\end{array}\right., \\[8mm]
c_2=(L_o-h_o/8\,\sin(\phi),0),\, 
\left\{\begin{array}{l}
X_2=h_o/4\,\sin(s+\pi/2) \\[2mm]
Y_2= \rho_o\,s 
\end{array}\right., \,
\end{array}
\ee
\end{itemize}
\par\noindent
\emph{3 Folds:} $L=2\,L_o$, $L_o=2.82\,W$, and $W=L_o/2\,\sin(\pi/4)$.
\par\noindent
\begin{itemize}
\item Circular arc, same folds; $\varphi=\pi/4.$,
\be\label{3F_arc_same}
\begin{array}{l}
c_1=(L_o/2,0), \quad 
c_2=(L_o,0), \quad
c_2=(3\,L_o/2,0), \quad
\left\{\begin{array}{l}
X_i = L_o/4\,\sin(s+\pi/2) \\[2mm]
Y_i = L_o/4\,\cos(s+\pi/2) 
\end{array}\right., 
\end{array}
\ee
\end{itemize}

%

\section*{Acknowledgments}  We gratefully acknowledge the support by the European Research Council through 
ERC PoC Grant Stripe-o-Morph (GA 101069436), 
by the EU H2020 program through I-Seed (GA 101017940) and Storm-Bots (GA 956150) projects, by the EU Horizon Europe program through MapWorms (GA 101046846) project
by the Italian Ministry of Research through the  projects Response (PRIN 2020), 
Abyss (PRIN 2022), and Innovative mathematical models for soft matter and hierarchical materials (PRIN 2022, PNRR F53D2300283 0006). The authors ADS and LT are members of the INdAM GNFM research group. ADS acknowledges useful discussions with Richard James, to whom this paper is dedicated, and with members of the research group Adaptive Beauty (J. Bico, B. Roman, T. Gao, N. Vani, T. Cheng, Y. Tahouni).


\bibliography{Biblio}

\begin{thebibliography}{10}
\expandafter\ifx\csname natexlab\endcsname\relax\def\natexlab#1{#1}\fi
\providecommand{\url}[1]{\texttt{#1}}
\providecommand{\href}[2]{#2}
\providecommand{\path}[1]{#1}
\providecommand{\DOIprefix}{doi:}
\providecommand{\ArXivprefix}{arXiv:}
\providecommand{\URLprefix}{URL: }
\providecommand{\Pubmedprefix}{pmid:}
\providecommand{\doi}[1]{\href{http://dx.doi.org/#1}{\path{#1}}}
\providecommand{\Pubmed}[1]{\href{pmid:#1}{\path{#1}}}
\providecommand{\bibinfo}[2]{#2}
\ifx\xfnm\relax \def\xfnm[#1]{\unskip,\space#1}\fi
\bibitem[{Alese(2022)}]{alese2022propagation}
\bibinfo{author}{Alese, L.}, \bibinfo{year}{2022}.
\newblock \bibinfo{title}{Propagation of curved folding: the folded annulus
  with multiple creases exists}.
\newblock \bibinfo{journal}{Beitr{\"a}ge zur Algebra und
  Geometrie/Contributions to Algebra and Geometry} \bibinfo{volume}{63},
  \bibinfo{pages}{19--43}.
\bibitem[{do~Carmo(1976)}]{doCarmo76}
\bibinfo{author}{do~Carmo, M.P.}, \bibinfo{year}{1976}.
\newblock \bibinfo{title}{Differential Geometry of curves and surfaces}.
\newblock \bibinfo{publisher}{Paperback}.
\bibitem[{Demaine et~al.(2018)Demaine, Demaine, Huffman, Koschitz and
  Tachi}]{demaine2018conic}
\bibinfo{author}{Demaine, E.D.}, \bibinfo{author}{Demaine, M.L.},
  \bibinfo{author}{Huffman, D.A.}, \bibinfo{author}{Koschitz, D.},
  \bibinfo{author}{Tachi, T.}, \bibinfo{year}{2018}.
\newblock \bibinfo{title}{Conic crease patterns with reflecting rule lines}.
\newblock \bibinfo{journal}{arXiv preprint arXiv:1812.01167} \bibinfo{note}{In
  Origami$^7$, Vol. 2. Tarquin, 574--590}.
\bibitem[{Feng et~al.(2024)Feng, Dradrach, Zmy{\'s}lony, Barnes and
  Biggins}]{feng2024geometry}
\bibinfo{author}{Feng, F.}, \bibinfo{author}{Dradrach, K.},
  \bibinfo{author}{Zmy{\'s}lony, M.}, \bibinfo{author}{Barnes, M.},
  \bibinfo{author}{Biggins, J.S.}, \bibinfo{year}{2024}.
\newblock \bibinfo{title}{Geometry, mechanics and actuation of intrinsically
  curved folds}.
\newblock \bibinfo{journal}{Soft Matter} \bibinfo{volume}{20},
  \bibinfo{pages}{2132--2140}.
\bibitem[{Fuchs and Tabachnikov(2006)}]{fuchs_omnibus}
\bibinfo{author}{Fuchs, D.}, \bibinfo{author}{Tabachnikov, S.},
  \bibinfo{year}{2006}.
\newblock \bibinfo{title}{Mathematical Omnibus: 30 lectures on classic
  mathematics}.
\newblock \bibinfo{publisher}{American Mathematical Society}.
\bibitem[{Gao et~al.(2020)Gao, Si{\'e}fert, DeSimone and Roman}]{gao2020shape}
\bibinfo{author}{Gao, T.}, \bibinfo{author}{Si{\'e}fert, E.},
  \bibinfo{author}{DeSimone, A.}, \bibinfo{author}{Roman, B.},
  \bibinfo{year}{2020}.
\newblock \bibinfo{title}{Shape programming by modulating actuation over
  hierarchical length scales}.
\newblock \bibinfo{journal}{Advanced Materials} \bibinfo{volume}{32},
  \bibinfo{pages}{2004515}.
\bibitem[{Jiang et~al.(2019)Jiang, Mundilova, Rist, Wallner and
  Pottmann}]{jiang2019curve}
\bibinfo{author}{Jiang, C.}, \bibinfo{author}{Mundilova, K.},
  \bibinfo{author}{Rist, F.}, \bibinfo{author}{Wallner, J.},
  \bibinfo{author}{Pottmann, H.}, \bibinfo{year}{2019}.
\newblock \bibinfo{title}{Curve-pleated structures}.
\newblock \bibinfo{journal}{ACM Transactions on Graphics (TOG)}
  \bibinfo{volume}{38}, \bibinfo{pages}{1--13}.
\bibitem[{Liu and James(2024)}]{liu2024design}
\bibinfo{author}{Liu, H.}, \bibinfo{author}{James, R.D.}, \bibinfo{year}{2024}.
\newblock \bibinfo{title}{Design of origami structures with curved tiles
  between the creases}.
\newblock \bibinfo{journal}{Journal of the Mechanics and Physics of Solids}
  \bibinfo{volume}{185}, \bibinfo{pages}{105559}.
\bibitem[{Tachi(2011)}]{tachi2011one}
\bibinfo{author}{Tachi, T.}, \bibinfo{year}{2011}.
\newblock \bibinfo{title}{One-dof rigid foldable structures from space curves},
  in: \bibinfo{booktitle}{Proceedings of the IABSE-IASS Symposium}, pp.
  \bibinfo{pages}{20--23}.
\bibitem[{Tahouni et~al.(2020)Tahouni, Cheng, Wood, Sachse, Thierer, Bischoff
  and Menges}]{tahouni2020self}
\bibinfo{author}{Tahouni, Y.}, \bibinfo{author}{Cheng, T.},
  \bibinfo{author}{Wood, D.}, \bibinfo{author}{Sachse, R.},
  \bibinfo{author}{Thierer, R.}, \bibinfo{author}{Bischoff, M.},
  \bibinfo{author}{Menges, A.}, \bibinfo{year}{2020}.
\newblock \bibinfo{title}{Self-shaping curved folding: A 4d-printing method for
  fabrication of self-folding curved crease structures}, in:
  \bibinfo{booktitle}{Proceedings of the 5th Annual ACM Symposium on
  Computational Fabrication}, pp. \bibinfo{pages}{1--11}.

\end{thebibliography}

\end{document}